\documentclass[twocolumn]{aastex631}

\usepackage[colorinlistoftodos]{todonotes}
\usepackage{amssymb}
\usepackage{comment}
\usepackage{multirow,acronym}
\usepackage{natbib}
\usepackage{makecell}
\usepackage{booktabs}
\usepackage{amsmath}
\usepackage{subfigure}
\usepackage{color}
\usepackage{xcolor}
\usepackage{hyperref}
\usepackage[T1]{fontenc}
\usepackage{graphicx}
\usepackage{bm}
\usepackage{CJK}

\hypersetup{colorlinks=true, citecolor=blue, 
linkcolor=cyan, urlcolor=magenta}

\usepackage{float}
\usepackage{amsmath,amssymb}
\usepackage{hyperref} 
\usepackage{graphicx} 
\usepackage{color}
\usepackage{verbatim}    
\usepackage{enumitem}
\usepackage{natbib}
\usepackage{multirow}

\newcommand{\proptosim}{\mathrel{\vcenter{
 \offinterlineskip\halign{\hfil$##$\cr
 \propto\cr\noalign{\kern2pt}\sim\cr\noalign{\kern-2pt}}}}}
\renewcommand{\b}[1]{\boldsymbol{#1}}

\newcommand{\response}[1]{{\bf\color{blue} #1}}

\renewcommand{\min}{\mathrm{min}}
\renewcommand{\max}{\mathrm{max}}
\newcommand{\au}{\mathrm{AU}}
\newcommand{\cm}{\mathrm{cm}}
\newcommand{\m}{\mathrm{m}}

\newcommand{\g}{\mathrm{g}}
\newcommand{\G}{\mathrm{G}}
\newcommand{\K}{\mathrm{K}}  
\newcommand{\km}{\mathrm{km}}

\newcommand{\pc}{\mathrm{pc}}

\newcommand{\s}{\mathrm{s}}
\newcommand{\yr}{\mathrm{yr}}

\renewcommand{\d}{\mathrm{d}}
\renewcommand{\m}{\mathrm{m}}
\newcommand{\f}{\mathrm{f}}
\newcommand{\e}{\mathrm{e}}

\newcommand{\p}{{\rm p}}

\renewcommand{\response}[1]{#1}

\newcommand{\DOA} {Department of Astronomy, School of
  Physics, Peking University, Beijing 100871, China}
\newcommand{\KIAA}{Kavli Institute for Astronomy and
  Astrophysics, Peking University, Beijing 100871, China}

\newcommand{\nbody}{$N$-body }
\newcommand{\rebound}{\texttt{REBOUND}}
\newcommand{\w}{{\rm w}}
\newcommand{\Myr}{{\rm Myr}}

\acrodef{OC}{open cluster}
\acrodef{BH}{black hole}
\acrodef{DF}{dynamic friction}
\acrodef{NDF}{negative dynamic friction}
\acrodef{IMF}{initial mass function}
\acrodef{AGN}{active galactic nucleus}

\begin{document}

\title{ Open Cluster Dynamics under the Influence of
  Outflow-Ambient Interactions}

\author[0009-0003-9343-8107]{Muxin Liu}
\affiliation{\DOA}
\affiliation{\KIAA}

\author[0000-0002-6540-7042]{Lile Wang}
\affiliation{\KIAA}
\affiliation{\DOA}

\author[0000-0002-6506-1985]{Xiaoting Fu} 
\affiliation{Purple Mountain
  Observatory, Chinese Academy of Sciences, Nanjing 210023,
  China}

\author[0000-0001-6947-5846]{Luis C. Ho}
\affiliation{\KIAA}
\affiliation{\DOA}

\correspondingauthor{Lile Wang} 
\email{lilew@pku.edu.cn}

\begin{abstract}
  \response{Stars with outflows impinging on} ambient gas experience
  accelerations due to the gravitational feedback from
  the interaction morphology between the outflow
  and the ambient gas. Such ``negative dynamical friction''
  (NDF), in contrast to the conventional ``dynamical
  friction'' (DF), is studied for its impact on the dynamics
  of open clusters (OCs) immersed in a uniform ambient
  gas. We modify the $N$-body integration code \rebound\
  with both NDF and DF implemented according to the outflow
  conditions of each star in a consistently
    constructed OC. The evolution of stars is also involved
  in determining the gas-star interactions throughout their
  stellar lives. Compared to DF-only and gas-free models
  with identical initial conditions, the NDF-affected
  cluster is puffier and evaporates faster, as indicated by
  various diagnostics, including lower velocity dispersions
  and larger half-mass and half-light radii. Neutron stars
  with fast winds are expelled from the cluster due to their
  intensive NDF effect, even without the ``kicks'' by
  asymmetric supernovae. Exploration of parameter space
  confirms that the NDF effect is generally enhanced with
  higher ambient gas densities, in qualitative agreement
  with the expression of acceleration.  Outflow-ambient
  interactions should be considered for the proper
  interpretation of the stellar dynamics
    evolution in clusters.
\end{abstract}


\section{Introduction}
\label{sec:intro}

\Acp{OC} offer valuable insights into the formation and
evolution of stars. Comprised of approximately $10^2$ to
$10^4$ stars that are gravitationally bound to each other
\citep{binney2011galactic}, OCs represent 
well-defined single stellar populations. They form from a
common molecular cloud collapse \citep{Krumholz}, with the
same metallicity and age.  As a result, \acp{OC} serve as
valuable tracers of star formation, chemical evolution,
kinematics, and gravitational dynamics \citep[see,
  e.g.,][]{Friel, 2007MNRAS.379...34M, 2011MNRAS.411.1495M,
    Krumholz, CG2022, fu2022, Magrini2023}. For instance,
they can be utilized to determine the metallicity of stellar
populations in different regions of the Milky Way, as their
distances can be constrained by their color-magnitude
diagrams \citep[see,
e.g.,][]{1953AJ.....58...61S}. Moreover, the kinematic
properties of \acp{OC} can be used to trace the rotation
curve of the Milky Way \citep[see, e.g.,][]{Tarricq}.
Thanks to the contributions of missions like Gaia
\citep{GAIA2016} in providing stellar astrometry
information, including precise position, parallax, and
proper motion, significant advancements have been made
\cite[e.g.,][]{GAIA2018, GAIA2021, GAIA2022}.  The member
stars of \acp{OC} and the properties of several thousand
clusters have been determined based on Gaia data \citep[see,
e.g.,][]{CG2022, Hunt2023}. Some OCs show tidal structures
and indicate interactions with their environments
\citep[see, e.g.,][]{Pang2022, Tarricq2022}.  These
advancements enable the dynamical characteristics study of
\acp{OC} through observations and facilitate comparison with
theoretical models.

Most Galactic \acp{OC} travel in the Milky Way thin disk on
near-circular orbits that pass through the
Galactic mid-plane several times in one orbital revolution
\citep[see, e.g.,][]{fu2022}. The impact of the detailed
interactions between the cluster member stars and the gas
clouds in the Galactic plane remains unclear. \response{Do the
interactions change the morphology of the star clusters \citep{2021A&A...647A.137J,2021ApJ...912..162P} and
lead to the loss of member stars?} Do the interactions differ
among different kinds of stars? These are still open
questions.

The ambient gas density likely will affect the dynamical characteristics of an OC as it travels through a gas cloud. A massive object moving through
a giant gas cloud experiences \ac{DF} \citep{Chandrasekhar,
  1999ApJ...513..252O, Edgar}. According to the standard
Bondi-Hoyle-Lyttleton accretion model
\citep{hoyle_lyttleton_1939, Bondi1944, Edgar}, it is
understood that due to the long-range nature of gravity, the
gas outside the Bondi radius is also influenced by gravity,
resulting in the formation of an overdense tail downstream
of the massive object. This overdense region exerts a
pulling force on the object in the opposite direction of its
motion, causing it to decelerate. This force is known as
\ac{DF} force. Works such as \citet{Kim_2007} and
\citet{Baruteau_2011} have shown that gas \ac{DF} can lead
to the hardening of binaries, and \citet{Tagawa_2020} have
pointed out that binaries can form through single-single
interactions by dissipating kinetic energy in a gaseous
medium. Other works, such as \citet{Tanaka_2002}, have shown
that \ac{DF} force can dampen the velocity dispersion of
\acp{BH} and stars, and \citet{Just} have demonstrated that
this dissipative force acting on stars in the disk can
result in an increased mass flow toward the supermassive BH
and an asymmetry in the phase space distribution
due to a rotating accretion disk.

However, the situation changes if stellar objects launch
powerful winds. \citet{2020MNRAS.492.2755G} demonstrated
that when the wind speed is sufficiently high, the extent of
the underdense region becomes substantial, resulting in an
overall gravitational force from the gas that aligns with
the object's motion. Consequently, the \ac{DF} becomes
negative in this scenario, causing the object to
accelerate. This phenomenon is known as \ac{NDF}. In other
related works, \citet{2020MNRAS.494.2327L} employed
hydrodynamic simulations to explore the changes in accretion
rate and the strength of \ac{NDF} in the presence of
outflows from compact objects. Additionally,
\citet{2022ApJ...932..108W} conducted a study on \ac{NDF} in
the context of a binary system using global 3D hydrodynamic
simulations. This work uses the \nbody integrator \rebound\
\citep{Rein} to study the \ac{NDF} impact on the dynamics of
\ac{OC}s immersed in a uniform ambient gas. Thanks to the
support of \rebound\ in allowing the incorporation of
additional physics, both \ac{NDF} and \ac{DF} are
implemented according to the outflow conditions of each star
in a consistently constructed \ac{OC}.

This paper is structured as follows. \S \ref{sec:method}
provides comprehensive descriptions of the gas-star
interactions incorporated, the stellar evolution model, and
the setup of the fiducial model. \S
\ref{sec:result-fiducial} analyzes and compares the impacts
of different types of gas-star interactions—namely,
``\ac{DF}'', ``\ac{NDF}'', and ``None''—on the dynamics and
stellar evolution of the fiducial model OC, both with and
without bulk motion. \S \ref{sec:result-var-model} explores
scenarios with various ambient gas densities. Discussion and
a summary are given in \S \ref{sec:summary}.

\section{Methods}\label{sec:method}

This work simulates the dynamics of gas-coupled OCs 
using \nbody simulations by adding modules to
\rebound. For accuracy, we adopt the \texttt{IAS15}
integrator \citep{2015MNRAS.446.1424R}, which is a 15th
order scheme with adaptive step size control. In order to
model the clusters properly, stellar particles are
created with proper mass distributions and evolved with
stellar evolution models that include compact
objects epochs along with the dynamics.

\subsection{Gas-star Interactions}\label{subsec:gas-star}

The interaction between gas and stars can generate either friction
or anti-friction, depending on the strength of the stellar
outflows. For stars that have no outflows, the analytic
approximations for DF are adopted from
\citet{1999ApJ...513..252O}:
\begin{equation}
  \label{eq:method-fric}
  a_{\rm DF} = \dfrac{4\pi G^2 \rho}{v_*^2} \times
  \begin{cases}
    & \ln \left[\Lambda\left(1-\dfrac{1}{\mathcal{M}^2}
      \right)^{1/2} \right]\ ,\ \mathcal{M} > 1\ \\ 
    & \dfrac{1}{2}\ln \left(\dfrac{ 1 + \mathcal{M}} { 1 -
      \mathcal{M} } \right) - \mathcal{M}\ ,\ \mathcal{M} <
      1\  ,
  \end{cases}
\end{equation}
where $v_*$ is the stellar velocity, $\rho$ is the
  ambient gas density, $\mathcal{M} \equiv v_*/c_s$ is the
Mach number in the gas, and $\Lambda \equiv b_\max / b_\min$
is the Coulomb factor. In this work, we assume
  that the $a_{\rm DF}$ direction is always aligned with
  the relative motion direction, because the
  timescales of hydrodynamic phenomena $\tau_{\rm hyd}$ are
  typically much shorter than the stellar orbital dynamic
  timescales $\tau_{\rm dyn}$,
\begin{equation}
  \label{eq:method-df-timescales}
  \begin{split}
    & \tau_{\rm hyd} \approx 0.02~\yr \times
      \left( \dfrac{l_{\rm hyd}}{R_\odot} \right)
      \left( \dfrac{v_*}{1~\km~\s^{-1}} \right)^{-1}\ ,
    \\
    & \tau_{\rm dyn} \approx 10^5~\yr \times
      \left( \dfrac{l_{\rm dyn}}{0.1~\pc} \right)
      \left( \dfrac{v_*}{1~\km~\s^{-1}} \right)^{-1}\ ,
  \end{split}
\end{equation}
in which $l_{\rm hyd}$ and $l_{\rm dyn}$ are the typical
spatial scale of the hydrodynamic process and the stellar
dynamics, respectively. The overdense structures causing
DF is typically a few times the stellar size
in absence of outflows (see e.g.,
\citealt{1999ApJ...513..252O}), and the typical orbital
scales of stars in OCs are no less than \response{about $ 10^{-1}~\pc$}
(see e.g., \citealt{2002A&A...383..153N})
(unless in close encounter events, which are typically rapid
and dwarf DF effects by the ambient
gas). Therefore, it is almost always eligible to assume that
the star-gas interaction structures settle at the steady or
quasi-steady states, and the direction of DF
stays along the stellar motion relative to the
gas.  We use the Bondi radius of each star,
$b_\min \equiv 2GM/c_s^2$, for the minimum impact parameter,
and we approximate $b_\max$ with the size upper limit of a
giant molecular cloud (\response{$100~\pc$} in this work; e.g.,
\citealt{1987ApJ...319..730S,
  2017ApJ...834...57M,2018ApJ...860..172S}).

Whenever a star has outflows, the gravitational feedback
coming from the interactions between such outflows and the
ambient gas will likely accelerate the star. We adopt the
analytic approximation described in 
\citet{2020MNRAS.492.2755G} and \citet{2020MNRAS.494.2327L},
\begin{equation}
  \label{eq:method-anti-fric}
  \begin{split}
    a_{\rm NDF}
    & = \pi G \rho \int_0^\pi \d\theta\ \cos\theta
      \sin\theta\ R_\s \\ 
    & \times \left\{ \dfrac{3}{2}\left[ 1 +
      \dfrac{2u(1-\cos \theta)}
      {R_\s^2\sin^2\theta / R_0^2}
      \right]^2
      - 2\left[ 1 + \dfrac{u^2}{R_\s^2/R_0^2} \right]
      \right\}\ ,
  \end{split}
\end{equation}
where $R_0 \equiv [\dot{m}_\w v_\w/(4\pi\rho v_*^2)]^{1/2}$
is the standoff distance \response{(where the total pressure
  of the incoming medium equals that of the outflow),
  $\dot{m}_\w$ is the outflow wind mass-loss rate, $v_\w$ is
  the wind radial velocity and $u\equiv v_*/v_\w$.
  $R_\s = R_0 [ 3 ( 1 - \theta \cot\theta) /
  \sin^2\theta]^{1/2}$ is the contact discontinuity location
  at the polar angle $\theta$ (where the gas over-densities
  are located) in spherical coordinates for which the star
  is at the origin, and the axis $\theta=0$ is in the
  opposite direction of $v_*$.} In general, for typical OCs,
$v_*$ spans $0.5-5~\km~\s^{-1}$ \citep[e.g.,][]
{Tarricq,2018A&A...619A.155S, 1984ApJ...284..643M}, which is
significantly lower than the $v_\w$ of almost all stellar
outflows. Even asymptotic giant branch (AGB) stars, which
are known for their slow outflows, still have
\response{$v_\w\gtrapprox 30~\km~\s^{-1}$}
\citep{2004agbs.book.....H, 2008A&A...487..645R}. Therefore,
one can take the $u \rightarrow 0$ limit of
eq.~\eqref{eq:method-anti-fric} and adopt a simple analytic
approximation, $a_{\rm NDF} \approx 8.18G \rho R_0$
\citep[see also][] {2020MNRAS.494.2327L}.

Similar to the DF case, the
  NDF direction stays along the
  direction of star to gas relative motion. While the
  stellar dynamic timescales are estimated similar to
  eq.~\eqref{eq:method-df-timescales}, the estimation of
  hydrodynamic timescales approximately reads,
\begin{equation}
  \label{eq:method-ndf-timescales}
  \tau_{\rm hyd} \approx 5~\yr \times
  \left( \dfrac{l_{\rm hyd}}{10^2~\au} \right)
  \left( \dfrac{v_{\rm w}}{10^2~\km~\s^{-1}}
  \right)^{-1}\ , 
\end{equation}
where the spatial scale $l_{\rm hyd}$ should stand for the
distance from the star to the bow shock, and $v_\w$ is the
stellar wind velocity. As one can easily observe, the
inequality $\tau_{\rm hyd} \ll \tau_{\rm dyn}$ still holds
in the NDF cases. 

There are multiple types of stars in an OC, and their
interactions with the ambient gas are different. Unless
otherwise noted, we use the anti-friction recipes for main
sequence stars, red giant branch (RGB) stars, and neutron
stars (NSs), while we assume that white dwarfs (WDs) and
\acp{BH} have {\it no} outflows and thus obey $a_{\rm DF}$
in eq.~\eqref{eq:method-fric}. Admittedly, WDs and BHs in
binaries can accrete and have consequent activities (e.g.,
disk winds, decretions, and jets), yet detailed discussions
on those phenomena only add to the complications and obscure
the effects concerned, and are beyond the scope of the
current paper. For main sequence stars and giants, the wind
properties are calibrated for stars with solar
metallicity. The mass-loss rates are approximated by Reimers
formula (e.g., \citealt{1978A&A....70..227K}; see also
\citealt{2016ApJ...823..102C}), \begin{equation}
  \label{eq:method-mdot-reimers}
  \begin{split}
  \dot{m}_\w & \approx 4 \times 10^{-13}~M_\odot~\yr^{-1}\times
  \left( \dfrac{L}{L_\odot} \right)
  \left( \dfrac{R}{R_\odot} \right)
  \left( \dfrac{M}{M_\odot} \right)^{-1}\\
    & \propto 4\times 10^{-13}~M_\odot~\yr^{-1}\times
      \left( \dfrac{M}{M_\odot} \right)^{3.3}\ ,
  \end{split}
\end{equation}
where we use the approximate scaling laws $L\propto M^{3.5}$ 
\citep{1938ApJ....88..472K} and $R\propto M^{0.8}$ 
\citep{1991Ap&SS.181..313D}, \response{with $L$ as the stellar luminosity, $R$ as the stellar radius, and $M$ as the stellar mass}. Such mass-loss rates are on the high
end of typical values, yet stars in an OC
are generally young and tend to have stronger
stellar winds.  The stellar wind velocity has significant
variations and uncertainties from star to star. We adopt a
simplified power-law fitting for main sequence stars,
\begin{equation}
  \label{eq:method-vw}
  v_\w\approx 400~\km~\s^{-1}\times
  \left( \dfrac{M}{M_\odot} \right)^{0.6}\ ,
\end{equation}
which yields observed values of $v_\w =
  400~\km~\s^{-1}$ for 
solar-mass stars \citep[e.g.,][]{2015NatCo...6.5947B}, 
and $v_\w \approx 2000~\km~\s^{-1}$ for type
O8 stars \citep[e.g.,][]{1989A&A...226..215B}.  The
uncertainties of wind properties are even greater for RGB
stars; for simplicity, we assume all of them have
$v_{\rm w,RGB} \approx 30~\km~\s^{-1}$ and
  $\dot{m}_{\rm w, RGB} \approx 3\times
  10^{-9}~M_\odot~\yr^{-1}$ \citep{2004agbs.book.....H, 
  2008A&A...487..645R}. NSs can have significant 
outflows due to complex acceleration
mechanisms, 
such as winds originating from young and intensely 
hot neutron stars are propelled by photons and neutrons near
the stellar surface
\citep[]{1981ApJ...251..311S,1986ApJ...309..141D}. We assume
$v_\w \approx 3\times 10^4~\km~\s^{-1} = 0.1 c$ ($c$ for
  the speed of light), and
  $\dot{m}_\w \approx 3\times 10^{-9}~M_\odot~\yr^{-1}$ 
\citep[e.g.,][]{1983PASJ...35...33K,1990ApJ...363..218P,
  2023MNRAS.518..623M}.

\subsection{Stellar Evolution Properties}
\label{sec:method-star-evo}

Stars in an OC evolve for hundreds of millions of
years before the cluster evaporates \citep[]{2010ARA&A..48..431P, Krumholz, 2020SSRv..216...64K}. 
Given that the main sequence lifetime of
an 8~$M_\odot$ star with solar metallicity is about 35 Myr
\citep[see stellar model \texttt{PARSEC v1.2s} \footnote{https://people.sissa.it/$\sim$sbressan/parsec.html};][]
{2012MNRAS.427..127B, 2014MNRAS.445.4287T,
  2015MNRAS.452.1068C},
 before the cluster evaporates all massive stars will have evolved off the main sequence.
Very massive stars (\response{$M\gtrapprox 25~M_\odot$}) are not considered 
in this work. We assume that each star has
finished its main sequence phase and entered its 
RGB stage at the time we consider 
significant gas-star interactions to occur. The evolution time of the RGB phase is
approximated by a piecewise function in line with 
the solar metallicity and solar composition stellar models 
(Z=0.017, Y=0.279) of \texttt{PARSEC v1.2s}:  
\begin{equation}
  \label{eq:method-t-rgb}
  t_{\rm RGB} = 
  \begin{cases}
    & 10^5~\yr\ ,\ M > 15~{M_\odot}\ \\ 
    & 3\times 10^8\times \left( \dfrac{M}{M_\odot} \right)^{-4}
       ~\yr\ ,\ M \leq 15~{M_\odot}\ .
  \end{cases}
\end{equation}
After the RGB stage, we assume that stars of $M< 8\;M_\odot$
will undergo their AGB phase and end up as WDs. Because the
AGB stage experiences massive outflows that lose \response{about $50\%$}
of the stellar mass within a relatively short period
\response{($\Delta t\lessapprox 10^5~\yr$)}, the impact of AGB outflows
due to anti-friction can be accounted as a pulse momentum
injection, \response{by integrating $\d v_*/\d t = a_{\rm NDF}$ assuming
$u\ll 1$ and
$\dot{m}_\w \approx \Delta M / \Delta t$. The
  increments regarding the velocity is then better estimated
  by the change in $v_*^2$, 
\begin{equation}
  \label{eq:method-dv-pulse}
  \begin{split}
     v_*\ \d v_* &= a_{\rm NDF}v_*\ \d t\ ;
    \\
     \int_0^{\Delta v^2}\dfrac{1}{2}\ \d v_*^2 &=\int_0^{\Delta t} a_{\rm NDF} v_*
      \ \d t \approx \int_0^{\Delta t} 8.18G\rho R_0 v_*\ \d t ;
    \\
     \Delta v^2 &\approx 8.18G\left(\dfrac{\rho v_\w\Delta M \Delta t}{\pi}\right)^{1/2}
    \\
     &\approx 3\times 10^{-2}~\km^2~\s^{-2}
      \times \left( \dfrac{\Delta M}{M_\odot} \right)^{1/2}
    \\
    &\times
      \left( \dfrac{v_\w}{30~\km~\s^{-1}} \right)^{1/2}
      \left( \dfrac{\Delta t}{10^5~\yr} \right)^{1/2}
    \\
    &\times
      \left( \dfrac{\rho}{30~m_p~\cm^{-3}}\right)^{1/2},
  \end{split}
\end{equation}
where $m_p$ is the proton mass. $\Delta M$ and $\Delta v^2$
are the total mass-loss and the increment of $v_*^2$ during
the period $\Delta t$, respectively.} We assume that the
consequent WD has $M_{\rm WD} = \min\{ 0.5 M, M_\odot \}$
(here $M$ is for the progenitor RGB star mass), and
$\Delta M$ is given by the decrease of mass before becoming
a WD.  For stars with $M > 8\;M_\odot$, we estimate their
$\Delta v^2$ similarly, assuming that the expansion period
of the supernova ejecta is $\Delta t = 10^4~\yr$,
$v_\w = 0.1c$, and $\Delta M$ is also deduced from
$M_{\rm NS} = \min\{ 0.1 M, 1.5 M_\odot \}$ and the
progenitor mass $M$. Note that we consider this pulse
momentum injection by AGB outflow in all types of gas-star
interactions and intentionally ignore the ``kicks'' in NS
(WD) momenta induced by the asymmetries of supernovae
(mass-loss in WD progenitor), so as not to obscure the
\ac{NDF} by such kicks (while \ac{NDF} only adds to the
effects of the kicks). Other epochs of stellar evolution are
ignored as they have negligible impact on the stellar
cluster dynamics.

\begin{deluxetable*}{cccccccc}
  \tablecolumns{2} \tabletypesize{\scriptsize}
  \tablewidth{500pt}
  \tablecaption{ Initial OC samples in this work }
  \label{tab:sims_samples}  
  \tablehead{ \colhead{Name} & \colhead{$N_{\rm cl}$} &
    \colhead{$M_{\rm cl}~[M_\odot]$} &
    \colhead{$R_{1/2}~[\pc]$} & \colhead{$W_0$} &
    \colhead{$\sigma_{v}~[\km~\s^{-1}]$} & \colhead{$F_{\rm
        b}~[\%]$} & \colhead{Description}
  \\
  \colhead{} & \colhead{(1)} & \colhead{(2)} & \colhead{(3)}
  & \colhead{(4)} & \colhead{(5)} & \colhead{(6)} &
  \colhead{} 
} 
  \startdata 
  S1500     & 1500 & 1193 & 2.0 & 5.0 & 0.40 & 0
  & Fiducial sample (\S \ref{sec:result-fiducial};
  \S \ref{sec:result-var-model}) \\
  S500      &  500 &  401 & 2.0 & 5.0 & 0.22 & 0
  & Reduced star number
  (\S \ref{sec:result-fiducial-sizes}) \\
  S3000     & 3000 & 2302 & 2.0 & 5.0 & 0.56 & 0
  & Increased star number
  (\S \ref{sec:result-fiducial-sizes}) \\
  S1500b    & 1500 & 1198 & 2.0 & 5.0 & 28.82& 100
  & All stars in binary systems
  (\S \ref{sec:result-fiducial-sizes}) \\
  S1500-seg & 1500 & 1121 & 2.0 & 5.0 & 0.33 & 0
  & Completely segregated sample
  (\S \ref{sec:cluster-segregation}) \\
  \enddata

  \tablecomments{All samples are in virial equilibrium
    \response{without the influence of a Galactic tidal
      field}. Except for S1500-seg, all samples are
    unsegregated. \\
    (1) Number of stars; (2) Total mass; (3)
    Half-mass radius; (4) Concentration parameter, (5)
    Velocity dispersion; (6) Fraction of binaries.
  }
\end{deluxetable*}

\begin{deluxetable*}{ccc}
  \tablecolumns{2} \tabletypesize{\scriptsize}
  \tablewidth{500pt}
  \tablecaption{ Physical parameters for \nbody simulations
    for OCs in different scenarios.}
  \label{tab:sims_list}  
  \tablehead{ \colhead{Series} & \colhead{Name} &
    \colhead{Description$^\dagger$} } 
  \startdata 
    & CNMF$^\dagger$ & Fiducial model (Cold Neutral Media;
    \S
    \ref{sec:result-fiducial-sizes} -
    \ref{sec:result-fiducial-ns}) 
    \\ 
  Fiducial & CNM3$^\dagger$ & $v_{\rm c} = 3~{\rm km\
    s}^{-1}$(\S \ref{sec:result-fiducial-bulk}) \\ 
  studies & CNM5$^\dagger$ & $v_{\rm c} = 5~{\rm km\
    s}^{-1}$(\S \ref{sec:result-fiducial-bulk}) \\ 
    & CNMF-tidal$^\dagger$ & Incorporating the Galactic
    potential (\S \ref{sec:galactic-potential})\\ 
  \hline
    & WNM0$^\dagger$ & $[\rho] = 0$, $T = 5000~{\rm K}$
  (Warm Neutral Media)\\
    & WNM1$^\dagger$ & $[\rho] = 1$, $T = 1000~{\rm K}$
  (Warm Neutral Media)\\
Various ambient & MD$^*$ & $[\rho] = 3$, $T = 30~{\rm K}$
  (Diffuse Molecular Regions)\\
gas models & M4$^*$ & $[\rho] = 4$, $T = 30~{\rm K}$
  (Molecular Clouds)\\
(\S \ref{sec:result-var-model}) & M5$^*$ & $[\rho] = 5$, $T
= 30~{\rm K}$ 
  (Molecular Clouds)\\
    & M7$^*$ & $[\rho] = 7$, $T = 30~{\rm K}$
  (Molecular Clouds)\\
    & AGND$^*$ & $[\rho] = 9$, $T = 30~{\rm K}$
  (AGN Disk Gases)\\
  \enddata

  \tablecomments{ All models presented here use the S1500
    initial sample (see Table \ref{tab:sims_samples}).
    \response{Except for CNMF-tidal, the external Galaxy’s
      potential is exclude.}  Only the properties different
    from the fiducial model are described;
    $[\rho]\equiv \log_{10} [\rho/(0.3~m_p~{\rm
      cm}^{-3})]$.\\
    $\dagger$: $\Delta t_{\rm evo} = 200~{\rm Myr}$.  $*$:
    $\Delta t_{\rm evo} = 20~{\rm Myr}$.  }
\end{deluxetable*}

\subsection{Setup of Fiducial and Other Models}
\label{sec:method-fiducial}

  The cluster samples used in the studies on open
  clusters are usually generated in initial virial
  equilibrium \response{that} can be achieved through various numerical
  methods \citep{1974A&A....37..183A}, such as N-body
  integration, the Monte Carlo method, and the fluid
  dynamical approach \citep[e.g.,][]{1973VA.....15...13A,
    1971ApJ...164..399S, 1971ApJ...166..483S,
    1970MNRAS.150...93L, 1970MNRAS.147..323L}. In this work,
  the initial sample is generated by \textsc{McLuster},
  which is an open-source tool that can set initial
  conditions for N-body simulations or generate artificial
  star clusters for direct study
  \citep{2011MNRAS.417.2300K}. The virial equilibrium
  cluster samples generated using \textsc{McLuster} are
  detailed in Table~\ref{tab:sims_samples}. All generated
  samples use the broken power-law initial mass function
  over the stellar mass range $0.2 < (M/M_\odot) < 25$
  \citep{1993MNRAS.262..545K,2002Sci...295...82K,2002AJ....124.2721R}.
\begin{equation}
  \label{eq:method-kroupa-imf}
  \xi(M) \propto
  \begin{cases}
    & M^{-2.35}\ ,\ M >0.5\, M_\odot\ \\
    & M^{-1.3}\ ,\  M <0.5\, M_\odot\ ,
  \end{cases}
\end{equation} 
with the $W_0$ parameter in the \citet{King1966} model,
specifying model concentration, set to 5.0, and the
half-mass radius along the line of sight set to
$2.0~\pc$.

\begin{figure} 
  \centering    
  \includegraphics[width=1.0\columnwidth]
  {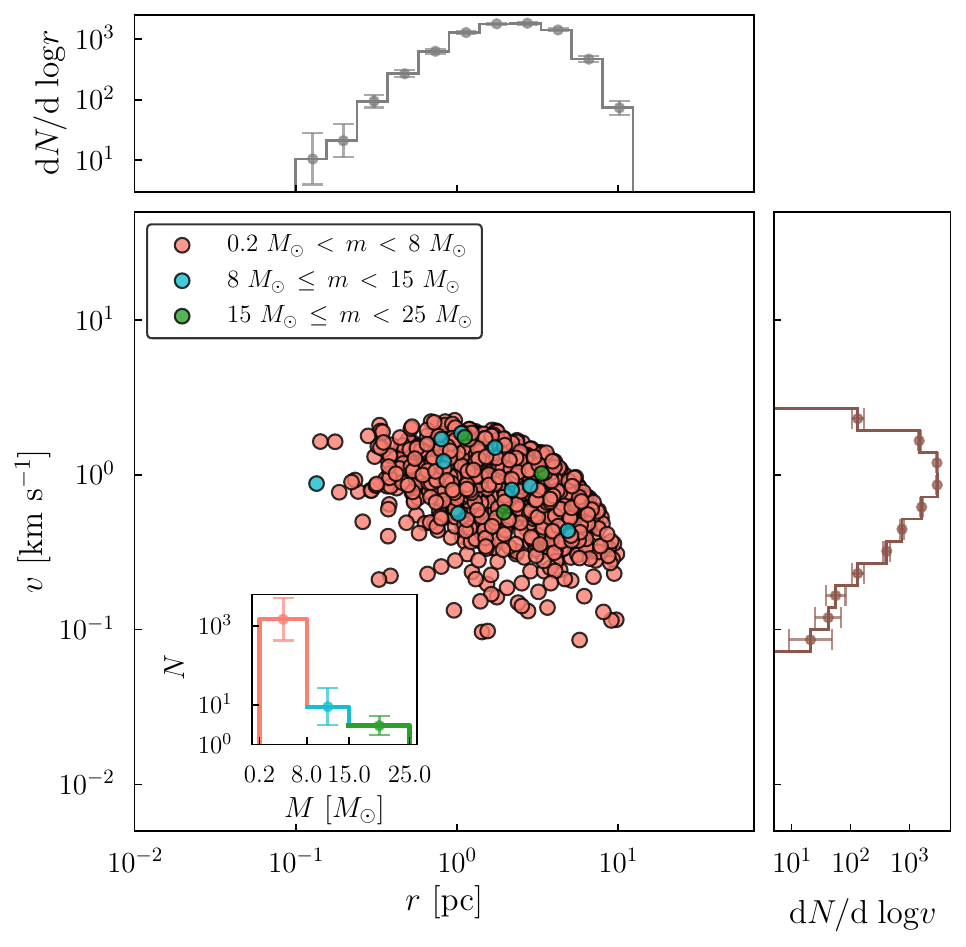}
  \caption{The initial condition of the fiducial
      sample S1500 includes logarithmic density histograms
    of distance (to the cluster center) and velocity in the
    upper and right panels, respectively. Mass bins are
    \response{indicated} by colors in the scattered plot. \response{The
      distance and velocity histograms} are calculated
      with respect to the centers of mass of bounded stars.
      The velocity dispersion of this sample is
      \response{ about $0.40~\km~\s^{-1}$}. }
  \label{fig:RVscatters-IC}     
\end{figure}

\begin{figure} 
  \centering    
  \includegraphics[width=1.0\columnwidth]
  {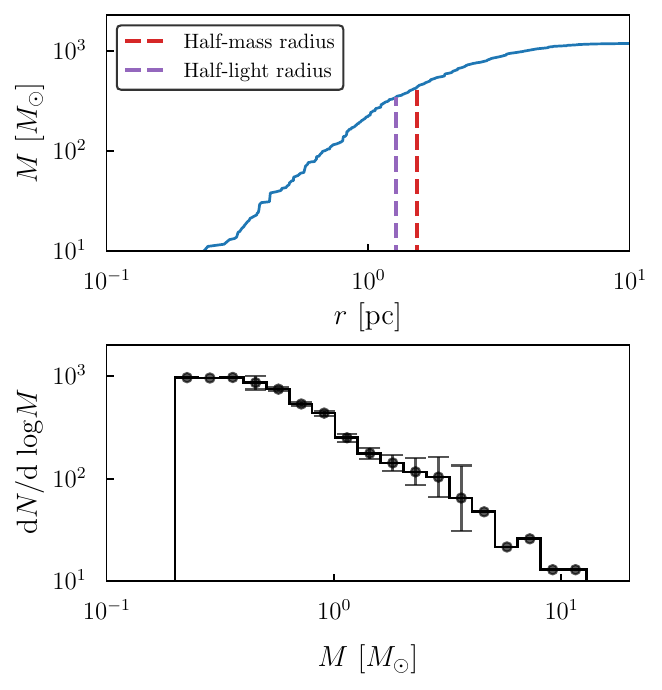}
  \caption{The initial condition of the fiducial
      sample S1500 includes the \response{stellar} mass distribution (top
      panel) and logarithmic density histograms of \response{stellar} mass
      (bottom panel\response{, the IMF}). The red and purple dashed lines in the
      top panel indicate half-mass ($1.54~\pc$) and
      half-light ($1.28~\pc$) radius, respectively.} 
  \label{fig:RVscatters-IC2}     
\end{figure}

\response{
With the help of \textsc{McLuster}, different initial 
conditions can be set when generating cluster samples by 
modifying the default parameter values. For example, 
``-Q'' sets the virial ratio to ensure the cluster is in 
virial equilibrium, ``-S'' adjusts the degree of mass 
segregation, and ``-b'' sets the binary fraction. The 
method used for generating samples with specified 
degrees of mass segregation and binary fraction is based 
on the approaches of \citet{2008ApJ...685..247B,1995MNRAS.277.1507K,
2008LNP...760..181K}.} The fiducial sample is a 1500-star ensemble S1500 without
segregation and binaries (see
Figures~\ref{fig:RVscatters-IC}--\ref{fig:RVscatters-IC2}),
which will be used for studying different physical scenarios
(\S 3 and 4). S500 and S3000 are two samples with different
numbers of stars, used for comparison with S1500 in the
fiducial model study. With the same stellar number as S1500,
S1500-seg is a fully mass-segregated cluster, where the most
massive star occupies the lowest energy orbit, and the least
massive star occupies the highest energy orbit. Note that
the previously mentioned samples do not include initial
binaries, as introducing binaries would complicate internal
interactions within the cluster. For example, collisions
could occur between the outflows of stars in binary systems
\citep[e.g.,][]{2022ApJ...932..108W}. Such complicated
interactions would require further study using hydrodynamic
simulations, which is beyond the scope of this current paper
and reserved for future works.

Following \citet{2001MNRAS.321..699K}, this work generates a
sample with a $100\%$ binary fraction through random pairing
(S1500b) to study simplified physical scenarios (considering
only gas-star interactions). The binary fraction is defined
as
\begin{equation}
  \label{eq:binary-fraction}
  F_{\rm b} \equiv \dfrac{N_{\rm bin}}{ N_{\rm bin} + N_{\rm single} }\ ,
\end{equation}
where $N_{\rm bin}$ is the number of binary and
$N_{\rm single}$ is the number of single stars. The method of
identifying binaries follows \citet{2009MNRAS.397.1577P},
which is calculating pairs with negative relative energy.

The fiducial model studies an OC immersed in the
  cold neutral medium (CNM), one of the phases of the
  interstellar medium (ISM) in the Galaxy, which has
  $\rho \approx 30~m_p~\cm^{-3}$ and $T \approx 100~\K$
  \citep{DraineBook}. The OC then evolved for
  $\Delta t_{\rm evo} = 200~\Myr$ within the bounding
  box. An important assumption is that star formation in the
  cluster is a single-age event, and once the cluster is
  formed, all the gas is expelled.  Different models
exploring physical parameters (especially those about the
ambient gas properties) use the same initial
fiducial sample S1500 (see
  Table~\ref{tab:sims_samples}-\ref{tab:sims_list}). Simulations
for each model, including the fiducial model specific to the
CNM, are conducted with three types of gas-star
interactions, indicated as ``NDF'', ``DF'' and ``None'' as
the suffixes of the models, to emphasize the modes of the
gas-star interactions. Note that the stars that do not
launch winds (e.g., WDs) always experience \response{the} DF effect even in
the model marked as ``NDF''.

\subsection{Basic Characterizations}
\label{sec:method-size}

Statistics about cluster sizes require a clear definition of
the cluster center. This work uses the center of mass of the
bounded stars, calculated through these two steps:
\begin{enumerate}
\item Select all stars within the $L_{\rm box} = 500~\pc$
  simulation bounding box and calculate their
  \response{total energy (the summation of kinetic and
    gravitational energy)} with respect to the center of
  mass. Stars with negative \response{total} energy are
  considered bounded.
\item Calculate the location and velocity of the center of
  mass for the bounded stars, which are used as the origin
  points in the configuration and momentum spaces in
  subsequent analyses.
\end{enumerate}
Similar to the center of mass, the ``center of light''
is also desired for the analyses of stellar luminosity
distribution, which is defined as the luminosity-weighted
average coordinates for the bounded stars identified in the
Step 1. Considering the observational conditions, unless
specifically emphasized, all half-mass and half-light radii
in this paper represent 2D values projected along the line
of sight, calculated by averaging over three lines of sight
as samples,
\begin{equation}
  \label{eq:projected-half-radii}
    r_{1/2} \equiv
    \dfrac{1}{3}\left(r_{1/2}^{xy}+r_{1/2}^{yz}+r_{1/2}^{zx}
      \right)\ , 
\end{equation}
where $\{r_{1/2}^{xy},r_{1/2}^{yz},r_{1/2}^{zx}\}$ are the
projected half-mass or half-light radii calculated on the
$x-y,\,y-z,\,z-x$ planes, respectively.

For better presentations of fluctuations in the cluster
characterizations, the uncertainties about statistical results
in this work are estimated by the non-parametric bootstrap
resampling method, which is a widely accepted and commonly
used statistical method in astrophysics \citep[see
also][]{10.1214/aos/1176344552,
  2012msma.book.....F,2023AJ....165...45M}. In detail, the bootstrap method generates datasets of the same length as the original data through resampling with replacement, repeated 1000 times. The standard deviations in each bin were then computed based on the results from these 1000 bootstrapped datasets.  

\section{Fiducial Model Results}
\label{sec:result-fiducial}

\begin{figure} 
  \centering    
  \includegraphics[width=1.0\columnwidth]
  {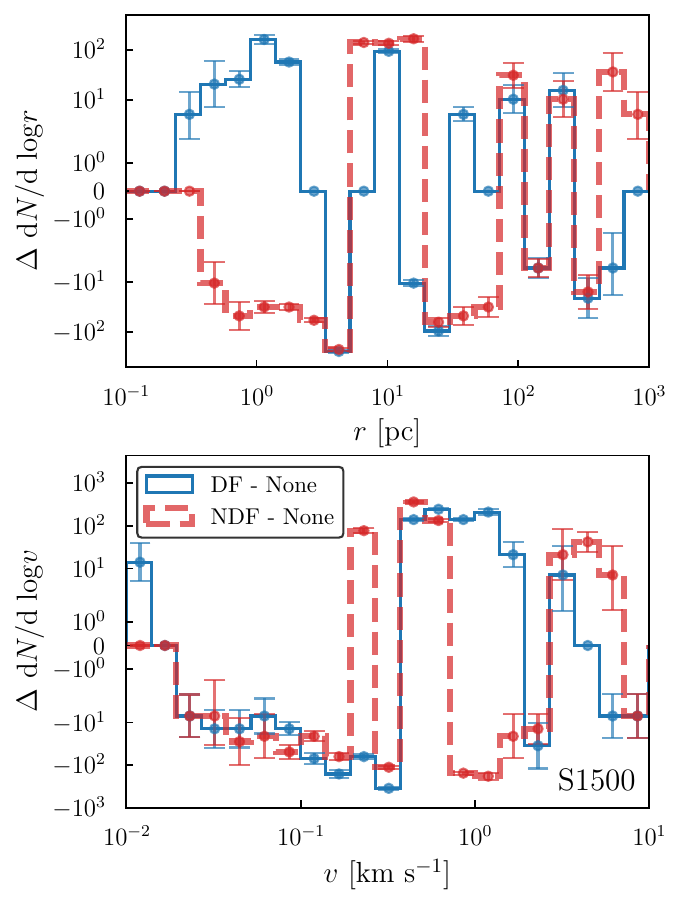}
  \caption{The impact of NDF and DF on the distance (upper
    panel) and velocity (lower panel) distributions of \response{stars in the} 
    OC in the CNMF \response{model}. Note that both panels display the differences
    compared to the None case in their distribution
    functions.}

  \label{fig:CNMF-RV}     
\end{figure}

\begin{figure*}
  \centering   
  \includegraphics[width=0.915\columnwidth]{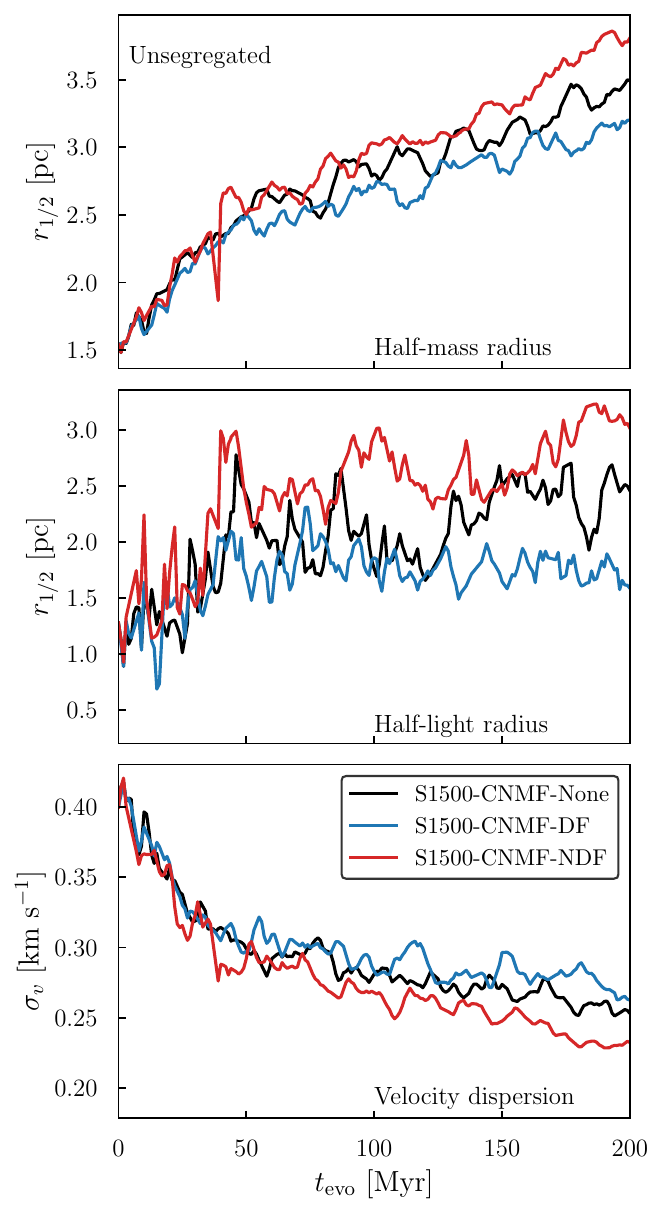}
  \includegraphics[width=0.908\columnwidth]
  {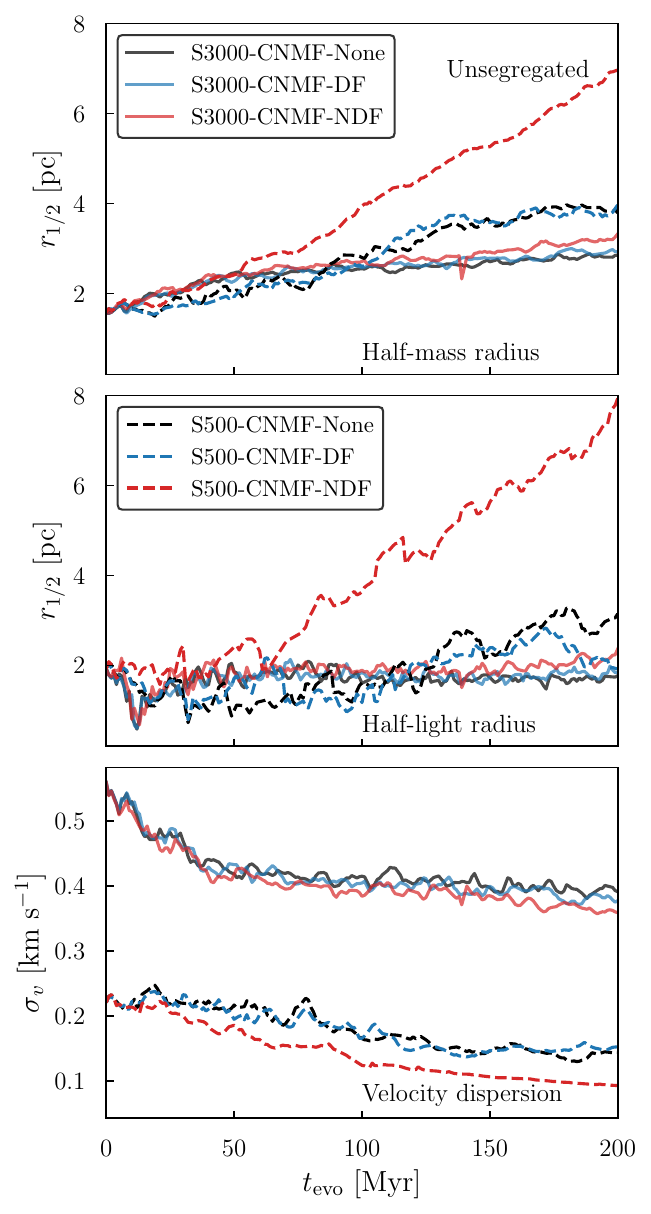}  
  \caption{The evolution of cluster properties under
    different gas-star interaction models. {\bf Left
      column}: the half-mass radii (upper panel), half-light
    radii (middle panel), and velocity dispersions (lower
    panel) of the \ac{OC} in the CNMF model during its
    $200~{\rm Myr}$ evolution period, with unsegregated
    initial sample S1500. {\bf Right column}: similar to the left
    column, showing the evolution of properties for models
    with different unsegregated initial samples, with line styles
    distinguishing the initial sample (solid for S3000,
    dashed for S500), and color for different star-gas
    interactions.}
  \label{fig:CNMF-unseg}
\end{figure*}

The impacts of the gas-star interactions on the model OC
immersed in the fiducial uniform CNM gas 
(CNMF hereafter for this case) are illustrated by
the histograms in Figures~\ref{fig:CNMF-RV}. Thanks to the
NDF acceleration, the stars generally reside on the side
with higher \response{total} energy in the CNMF-NDF run, compared
to its CNMF-DF and CNMF-None counterparts.  The distribution
of distances to the cluster center (``distances'' hereafter)
tilts to the more distant side, while the velocity tends to
be smaller as the stars are statistically farther from the
global gravitational potential minimum.

For a sun-like star \response{orbiting the Galaxy} embedded in the CNM gas,
  the velocity changes caused by NDF are as,
\begin{equation}
  \label{eq:solar-like-v-ndf}
  \begin{split}
    \eta_{v_*} & \equiv \dfrac{\Delta v_*}{v_*} =
                 \dfrac{a_{\rm NDF}\Delta t}{v_*} \approx
                 0.1\%\times 
          \left( \dfrac{v_\w}{400~\km~\s^{-1}} \right)^{1/4}
     \\
        &  \times \left(\dfrac{\rho}{30~m_p~\cm^{-3}} \right)^{1/4}
           \left(\dfrac{\dot{m}_\w}{4\times10^{-13}~M_\odot~\yr^{-1}}
          \right)^{1/4}  
     \\
        &  \times \left( \dfrac{\Delta t}{100~\Myr} \right)^{1/2}
           \left(\dfrac{v_*}{26~\km~\s^{-1}}\right)^{-1}\ ,
  \end{split}
\end{equation}
where $\Delta v_*$ is the stellar velocity variation caused
by NDF. Based on the conservation of angular momentum and
assuming the star moves along a circular orbit, the orbital
radius of the solar-like star will decrease by
\begin{equation}
  \label{eq:solar-like-r-ndf}
  \begin{split}
    \eta_{r_{\rm o}}
    & \equiv \dfrac{|\Delta r_{\rm
      o}|}{r_{\rm o}}
      = 1-\dfrac{v_*}{v_*+\Delta v_*} 
      = 1-\dfrac{1}{1+\eta_{v_*}} 
    \\
        & \approx 0.1\%\ , 
  \end{split}
\end{equation}
where $r_{\rm o}$ is the orbital radius and
$\Delta r_{\rm o}$ is the orbital variation caused by
NDF. The above calculations show that after $100~\Myr$ of
evolution, the orbit and velocity change by only $0.1\%$,
which is insignificant. Additionally, the gas component that
occupies the largest volume in the galaxy's ISM is the warm
neutral medium (WNM), which has a lower density than the CNM
(see \S \ref{sec:result-var-model} for different gas density
cases). This implies that the influence of NDF will be even
smaller. Hence, for a solar-like star embedded in the ISM
orbiting the galaxy, NDF will not significantly affect its
orbital motion unless it interacts with higher density gas.

\subsection{Cluster Sizes and Velocity Dispersions}
\label{sec:result-fiducial-sizes}

Statistics directly related to observables are presented in
the left column of Figure~\ref{fig:CNMF-unseg}. All three
models exhibit the same trend of dynamical evaporation,
indicated by their increasing half-mass and half-light radii
(in each time interval without star deaths), and decreasing
velocity dispersions. The half-mass radii for the
  CNMF-NDF model are larger than the gas-free CNMF-None
  model by \response{about $15\%$}, while those for the CNMF-DF model
  are smaller by approximately the same
  percentage. Acceleration forces (NDF) tend to increase the
  \response{total} energy of stars, relocating them to shallower
  regions in the potential well and reducing the magnitude
  \response{of the velocity dispersion}.  The deceleration forces (DF), in
  contrast, tend to result in larger velocity dispersions
  and smaller cluster sizes.

  One can observe ``cliffs'' at
  \response{$t_{\rm evo}\approx 39~\Myr$} in both the half-mass and
  half-light radii for the unsegregated NDF model in
  Figure~\ref{fig:CNMF-unseg}. These cliffs are the
  consequences of both the expulsion of massive stars
  \response{($M\gtrapprox 7~M_\odot$)} due to NDF acceleration , and the
  termination of massive stars on the main sequence as
  well. They are more pronounced in the half-light radius,
  as the luminosity is more sensitive to stellar mass
  ($L\propto M^{3.5}$).  In contrast, the DF models slow
  down the massive stars more efficiently, delaying their
  evaporation and resulting in a slower increase in cluster
  size compared to the case without gas-star interaction.

  To illustrate the influence of the star numbers,
  the results for samples S500 and S3000 with different
  numbers of stars are presented in the right column of
  Figure~\ref{fig:CNMF-unseg}. The increasing cluster size
  and decreasing velocity dispersion over time are
  qualitatively similar to S1500, while the cluster size
  grows much faster with NDF in S500 and slower in S3000.
  With decreasing star numbers, the evaporation timescale
  comes closer to the crossing timescale
  ($t_{\rm evap}\propto N_{\rm cl}t_{\rm cross}/\ln N_{\rm
    cl}$; see also \citealt{binney2011galactic}), making the
  cluster more susceptible to the acceleration caused by
  NDF, and vice versa.  
  
\begin{figure} 
    \centering    
    \includegraphics[width=1.0\columnwidth]
    {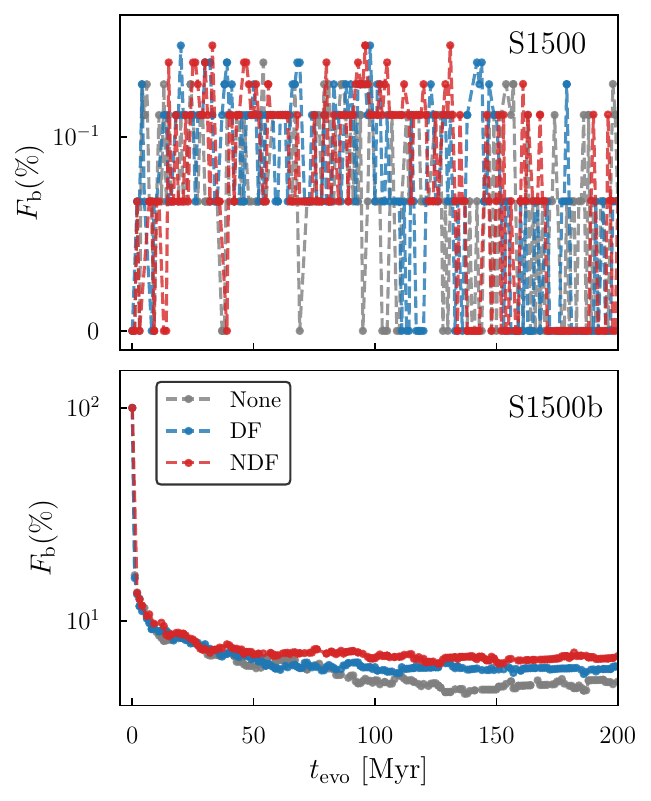}
    \caption{The evolution of binary fraction of
        \ac{OC} for all runs in the CNMF model using S1500
        and S1500b.}
    \label{fig:binary_3e1-100}     
\end{figure}

  The formation and evolution of binaries within
  \ac{OC}s has long been a topic of concern
  \citep[e.g.,][]{1995MNRAS.277.1491K,
    1995MNRAS.277.1507K,1995MNRAS.277.1522K,
    2018ApJ...859...36L}. Figure~\ref{fig:binary_3e1-100}
  illustrates the evolution of the binary fraction. For both
  S1500 and S1500b, the three runs for the CNMF model do not
  exhibit significant differences in $F_{\rm b}$. For
  S1500b, binary systems are rapidly disrupted in the first
  1 Myr, during which the soft binaries are
  destroyed. Subsequently, the clusters reach an equilibrium
  where the binary fraction becomes roughly stable. This
  behavior is consistent with \citet{1995MNRAS.277.1491K,
    1995MNRAS.277.1507K, 1999NewA....4..495K,
    2009MNRAS.397.1577P}.

\subsection{Mass-segregated Clusters}
\label{sec:cluster-segregation}

\begin{figure} 
  \centering    
  \includegraphics[width=1.0\columnwidth]
  {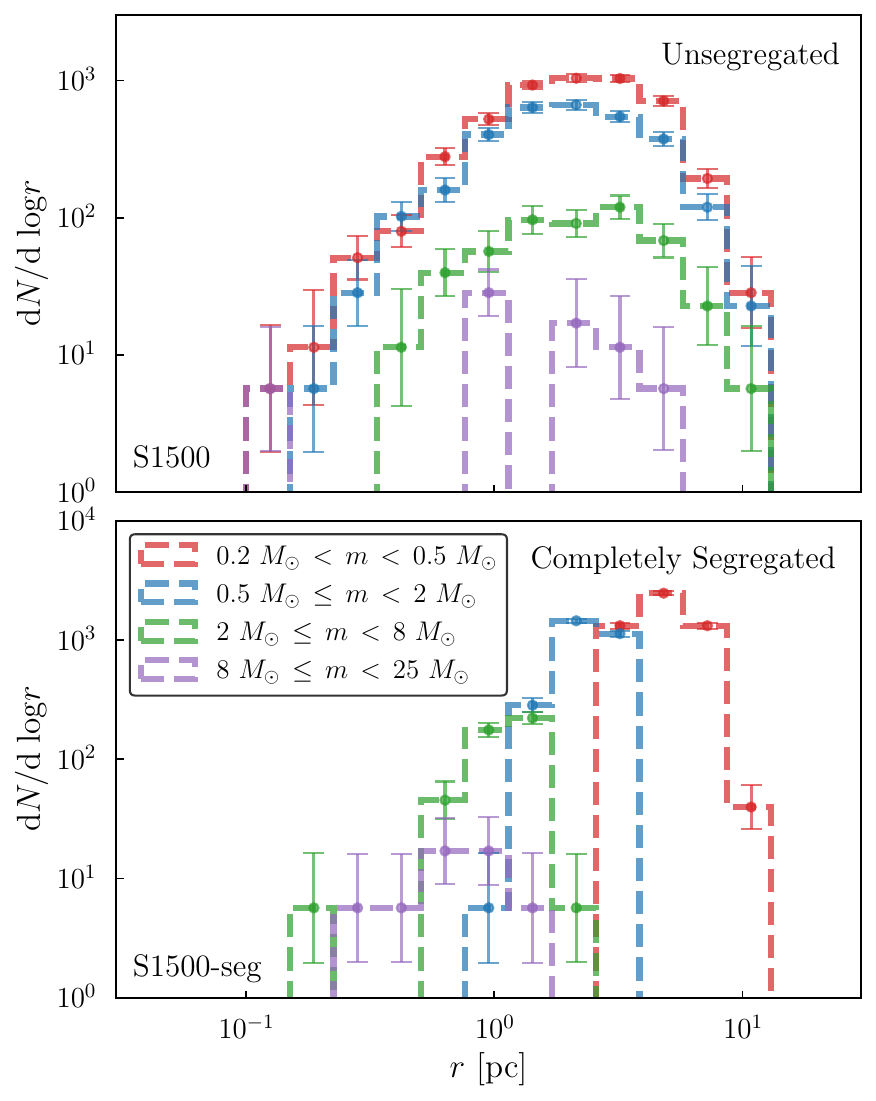}
  \caption{The distance distribution for different
      mass bins in the unsegregated (S1500) and completely
      segregated samples (S1500-seg).}
  \label{fig:mcluster-seg}     
\end{figure}

\begin{figure} 
  \centering    
  \includegraphics[width=0.9\columnwidth]
  {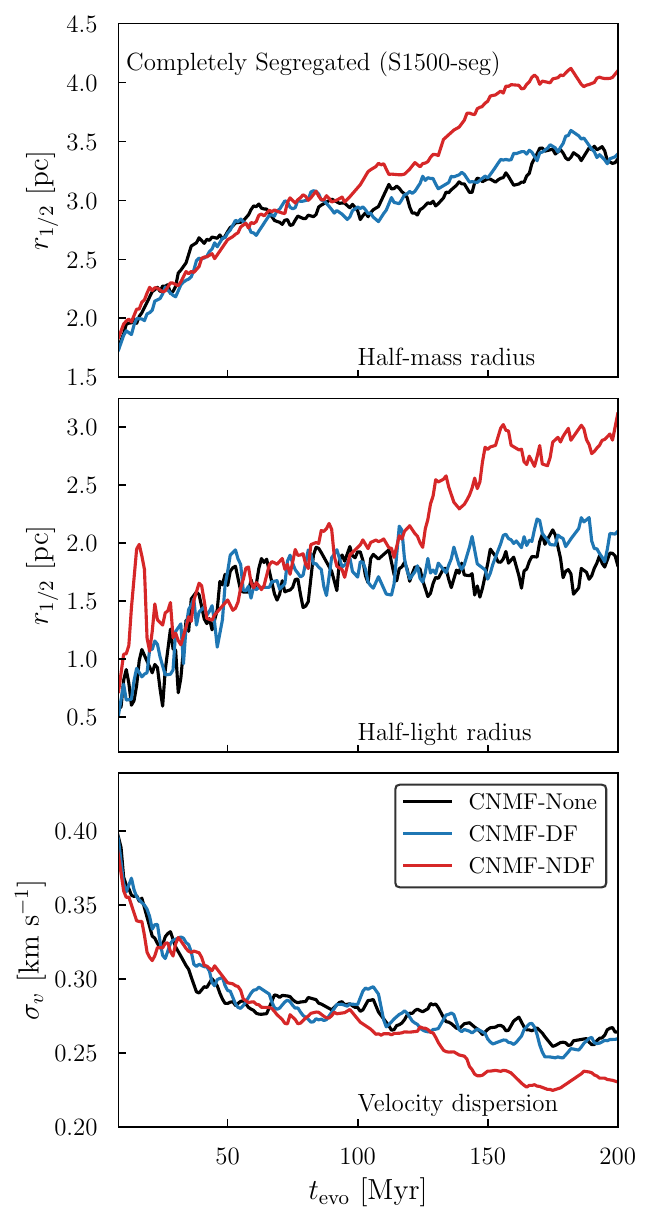}      
  \caption{Similar to the left column of
    Figure~\ref{fig:CNMF-unseg}, but
    showing the cluster properties for the completely
    segregated models. }
  \label{fig:CNMF-seg}     
\end{figure}

Actual stellar clusters may have different \response{extents} of mass
segregation during their formation and early evolution, that
more massive stars tend to \response{be distributed} closer to the cluster
center \citep[e.g.][] {Chandrasekhar,
  Chandrasekhar_Neumann}. We study a completely
  segregated cluster sample S1500-seg generated by McLuster
  (see also Table \ref{tab:sims_samples} and
  Figure~\ref{fig:mcluster-seg}), and present the evolution
  of properties in Figure~\ref{fig:CNMF-seg}.

  The differences in the half-light \response{radii} and half-mass
  radii between NDF and the other two models are more
  significant than \response{for the fiducial model S1500}. 
  After $200~\Myr$ evolution, the half-light radii
  for the NDF model are about twice as large as in the DF
  and gas-free models. The NDF model ``overturns'' the
  segregation by raising more massive stars to larger
  distances due to stronger acceleration, while the DF and
  gas-free models only add to the segregation during the
  evolution. With more numerical experiments not presented
  here, we confirm the general trend that the more
  mass-segregated the clusters are initially, the more
  important the NDF becomes along the dynamical evolution of
  OCs.

\subsection{Bulk Motion of Clusters}
\label{sec:result-fiducial-bulk}

\begin{figure} 
  \centering    
  \includegraphics[width=1.0\columnwidth]
  {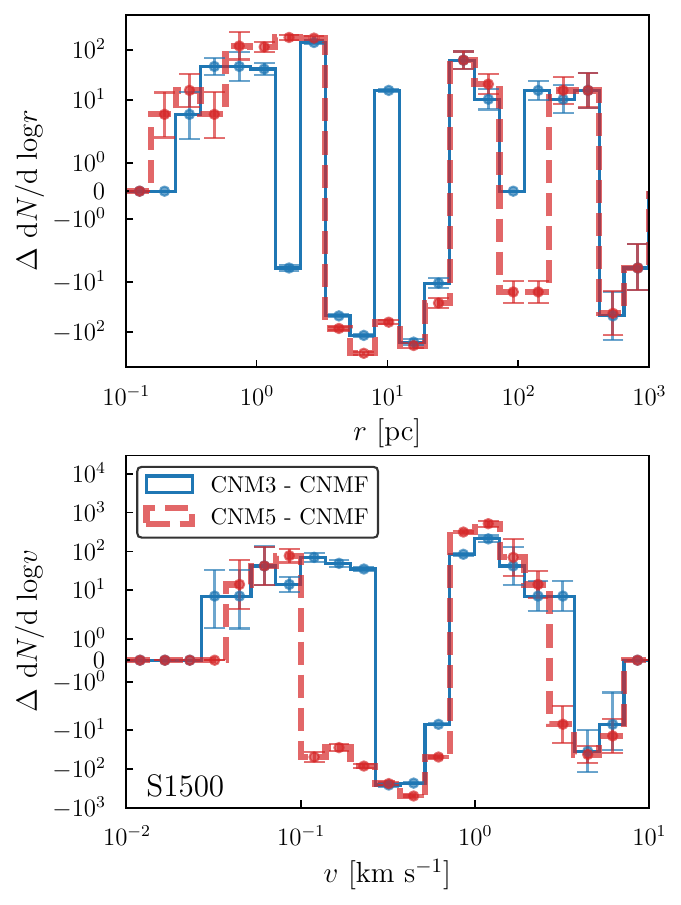}
  \caption{The impact of NDF on the distance (upper panel)
    and velocity (lower panel) distributions of \ac{OC} with
    different bulk motion speed $v_{\rm c}$. Note that both
    panels display the differences compared to CNMF-NDF in
    their distribution functions.}
  \label{fig:CNMF-3-5-RV}     
\end{figure}

The gas-star interactions--both NDF and DF--are
the results of stellar outflows and motion. Although the
CNMF model runs assume no bulk motion of the whole cluster,
the \response{real situation is} usually more complicated. The centers of
mass of an OC and a gas cloud can move on their own tracks
when evolving in galaxies, and the relative bulk motion
between the two objects should be expected. We hence
introduce models CNM3 and CNM5 to study such bulk motions at
center of mass speeds $v_{\rm c} = 3~\km~\s^{-1}$ and
$5~\km~\s^{-1}$, respectively (see also
Table~\ref{tab:sims_list}).

\begin{figure} 
  \centering    
  \includegraphics[width=1.0\columnwidth]
  {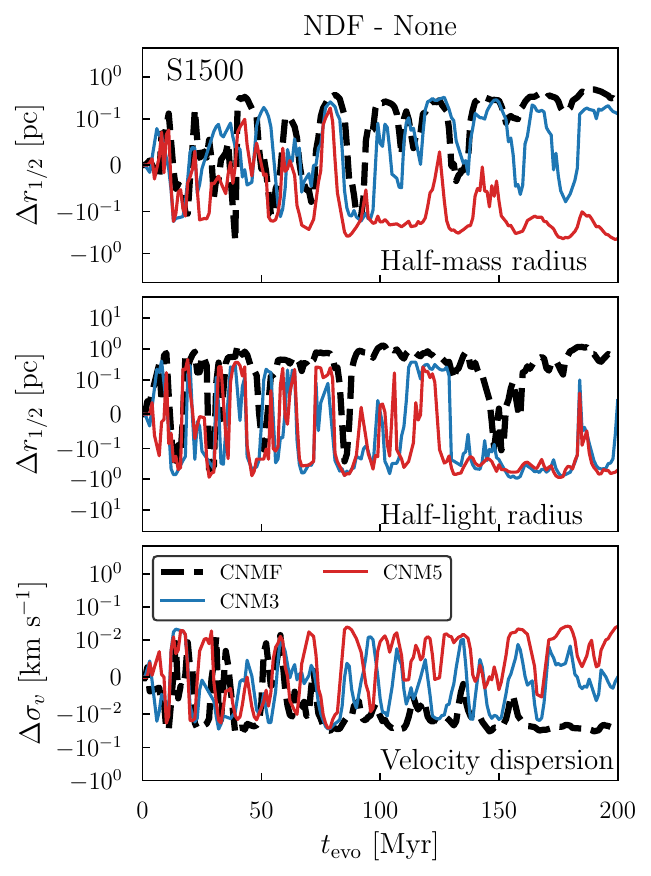}
  \caption{The evolution of differences in half-mass radii
    (upper panel), half-light radii (middle panel), and
    velocity dispersions (lower panel) across different
    scenarios (CNMF, CNM3 and CNM5). The
      differences are obtained by subtracting the values of
      the gas-free case (``None'') from the NDF case). } 
  \label{fig:Radius-NDF-DF-2}     
\end{figure}

Due to the morphologies of the stellar wind-ambient
interaction regions, the magnitudes of NDF acceleration
decreases with faster stellar velocities
(eq.~\ref{eq:method-anti-fric}). This relation qualitatively
yields the phenomena in Figure~\ref{fig:CNMF-3-5-RV}, where
the increase in \response{total} energy caused by NDF is
slightly suppressed with higher $v_{\rm c}$. Regarding the
increase in half-mass and half-light radii, a similar
suppression is observed in
Figure~\ref{fig:Radius-NDF-DF-2}. Concerning velocity
dispersion, the NDF suppression results in stars settling in
deeper regions of the potential well, leading to larger
velocity dispersion than \response{in} the CNMF
\response{model}. While these patterns can be qualitatively
anticipated for higher $v_{\rm c}$ velocities, detailed
analyses should be based on cluster-specific calculations
rather than extrapolations.

\subsection{Stellar Evolution and Compact Objects}
\label{sec:result-fiducial-ns}

\begin{figure} 
  \centering    
  \includegraphics[width=1.0\columnwidth]
  {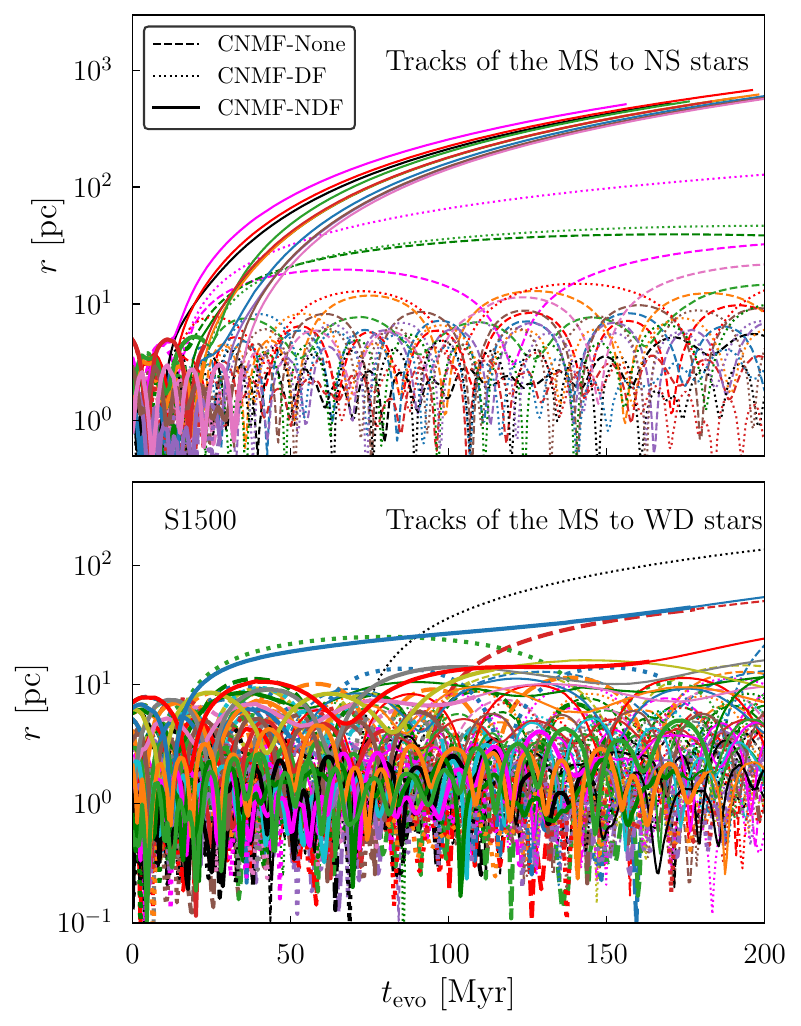}
  \caption{The evolution trajectories of stars evolving into
    compact objects (NS in the upper panel, and WD in the
    lower panel), showing the distances $r$ to the cluster
    center, in \response{the CNMF model} with distinct stars
    represented by different colors. The upper and lower
    panels show the tracks of main sequence stars
    transitioning into NS and WDs under different gas-star
    interactions. Thick lines represent trajectories during
    the main sequence phase, while thin lines indicate
    trajectories after they evolve into NS or WD. These
    trajectories only depict stars exiting the main sequence
    within the evolution time. Note that all NSs are
    expelled from \response{the} OC under the influence of
    NDF.}
  \label{fig:track_3e1-100}     
\end{figure}

\begin{figure} 
    \centering    
    \includegraphics[width=1.0\columnwidth]
    {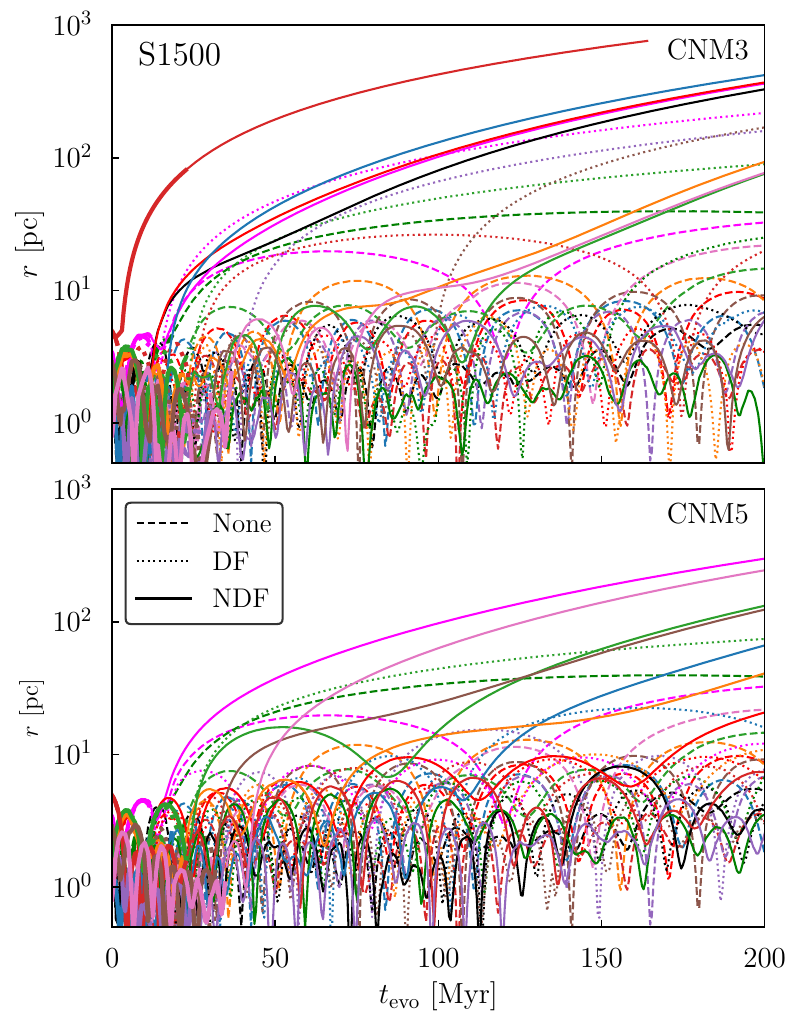}
    \caption{The evolutionary trajectories of main sequence
      stars transitioning into NSs are depicted in the
      scenarios of CNM3 (upper panel) and CNM5 (lower
      panel). In both CNM3 and CNM5, some NSs remain within
      the \ac{OC} but tend to move outward. This phenomenon
      is attributed to the bulk motion attenuating the NDF
      effect.}
    \label{fig:track_CNM3_CNM5}     
\end{figure}

Outflowing stars are expected to be pushed further away from
the cluster center once NDF is in action. For typical main
sequence stars immersed in the CNM, however, such effects
are generally obscured by the overall trend of OC dynamical
evaporation. When an outflowing star travels at a low
initial speed in a uniform medium, the asymptotic relation
near $t\rightarrow \infty$ between the travel distance $l$
and the time $t$ roughly reads, \response{using the
  approximate expression for eq.~\eqref{eq:method-anti-fric}
  and integrating $\d^2 l/\d t^2 \approx a_{\rm NDF}$, }
\begin{equation}
  \label{eq:result-expel-free}
  \begin{split}  
    l & \approx 7~\pc \times \left(
        \dfrac{\rho}{30~m_p~\cm^{-3}} \right)^{1/4}
        \left( \dfrac{v_\w}{400~\km~\s^{-1}} \right)^{1/4}
    \\
      & \ \times \left(
        \dfrac{\dot{m}_\w}{10^{-13}~M_\odot~\yr^{-1}}
        \right)^{1/4} \left(\dfrac{t}{100~\Myr}\right)^{3/2}
        \propto t^{3/2}\ .
  \end{split}
\end{equation}
In the meantime, the velocity scales as $v\propto t^{1/2}$
asymptotically. These estimations also give the {\it lower}
limit of time required to drive a star out of the cluster
even {\it without} the gravitational potential well. When
the potential well is present, direct exclusion of a
solar-mass main sequence star only via NDF is not possible.

For objects with intensive outflows, however, the
accelerations are much more significant dynamically.
Although $a_{\rm NDF}\propto v_\w^{1/2}$ increases
sub-linearly with $v_\w$, the NS with $v_\w = 0.1 c$ can
still escape from the stellar ensemble well before the
evaporation of OCs. Figure~\ref{fig:track_3e1-100}
illustrates the efficient expulsion of NS (upper panel) by
NDF (solid lines), compared to the DF (dotted lines) and
gas-free (dashed lines) cases where the ordinary
``evaporation'' process of \response{an} OC and the increment of \response{a} star's
velocity by AGB-outflow-induced pulse \response{are} present. NS are driven
out to \response{$\gtrapprox 10^2~\pc$} from the center of the cluster
within only \response{about $50~\Myr$}, shorter than the lifetime of a
typical OC. After escaping from the OC potential, the
$l\propto t^{3/2}$ scaling and
eq.~\eqref{eq:result-expel-free} holds semi-quantitatively
for each of the NS at large distances \response{($\gtrapprox
10~\pc$)}. Assuming that the gaseous disk half-thickness is
\response{about $0.5~{\rm kpc}$} in the Galaxy, such expulsion process
will eventually take \response{about $100~\Myr$} to remove an NS from
the disk. Such quick expulsions will likely lead to the
scarcity of NS in OCs, and probably in the whole gaseous
galactic disk. Figure~\ref{fig:track_CNM3_CNM5} illustrates
a reduced expulsion of NSs attributed to the attenuation of
the NDF effect by the bulk motion. The NS expulsion is
ubiquitous for gas-immersed OCs unless the cluster bulk
motion is too fast relative to the gas. Previous works often
attribute the NS exclusion from clusters to the ``kicks''
during asymmetric supernovae explosions \citep[e.g.,][]
{2020ApJ...901L..16F}. With the NDF effects, however, NS
exclusion can still take place without these ``kicks''.
Note that even in the case of intensive outflows from stars,
the winds of the stars cannot combine to drive a wind from
the OC. This is because the standoff distance of the contact
discontinuity,
\begin{equation}
  \label{eq:result-standoff-distances}
  \begin{split}  
    & R_0  \approx 0.004~\pc \times
      \left( \dfrac{v_\w}{400~\km~\s^{-1}} \right)^{1/2}
      \left(\dfrac{v_*}{1~\km~\s^{-1}}\right)^{-1}
    \\
    & \ \times \left(
      \dfrac{\dot{m}_\w}{4\times10^{-13}~M_\odot~\yr^{-1}}
      \right)^{1/2}
      \left(\dfrac{\rho}{30~m_p~\cm^{-3}} \right)^{-1/2}\ ,
  \end{split}
\end{equation}
is much smaller than the distance between member stars
(e.g., the minimum distance in the initial OC of the
fiducial case is \response{about $0.01~\pc$}).

\begin{figure} 
  \centering    
  \includegraphics[width=1.0\columnwidth]
  {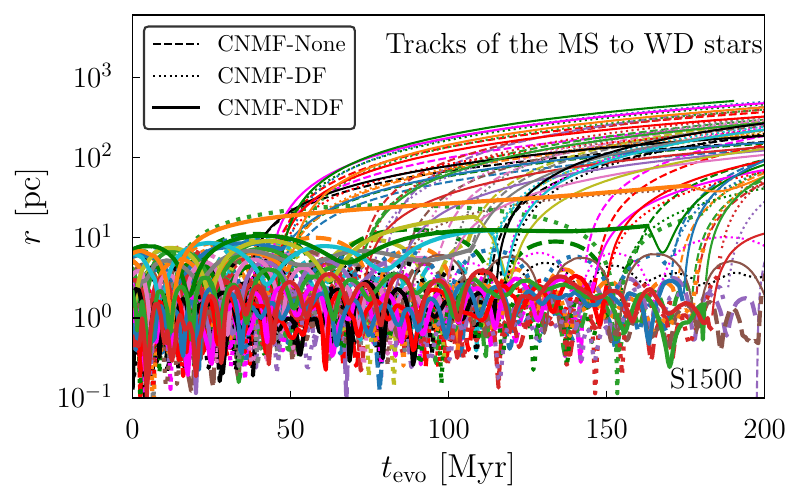}
  \caption{The evolutionary trajectories of main sequence
    stars transitioning into WDs. Note that in this
    scenario, the pulse is set with a direction that
    uniformly distributes randomly in space and a magnitude
    of $2~\km~\s^{-1}$.}
  \label{fig:wd_track_3e1-100n}     
\end{figure}

  In contrast to NS, WDs do not launch outflows in
  our models.  This causes the absence of NDF during their
  WD \response{epochs}, and the pulsive acceleration in their AGB
  stages is clearly insufficient to exclude them from the
  OCs.  However, the deficiency of WDs in young OCs has been
  proved by observations
  \citep[e.g.][]{2001AJ....122.3239K}. Once the WD outflows
  due to binary accretion processes are included (which
  requires detailed modeling in feedback and the
  non-isotropic outflow patterns,
  e.g. \citealt{2020MNRAS.494.2327L}, and is also postponed
  to future works), WDs are also expected to experience
  expulsion in their \response{dynamical} evolution. To illustrate
  another possible exclusion mechanism of WDs, we set up a
  model that enables pulses in velocity for newly formed WDs
  with a magnitude of $2~\km~\s^{-1}$ and random directions,
  following \citet{2003ApJ...595L..53F}. As one can see from
  Figure~\ref{fig:wd_track_3e1-100n}, such pulses can indeed
  cause the exclusion of WDs regardless to the condition of
  gas-star interactions.

\subsection{Cluster in Galactic Potential}
\label{sec:galactic-potential}

As a star departs from its cluster, its path is computed
considering the Galaxy's external potential. Incorporating
the Galactic potential in models of OC offers a more
realistic scenario. Following \citet{1991RMxAA..22..255A},
the Galactic potential $\phi_{\rm Gal}$ is composed of three
parts: the central part $\phi_1$, the disc $\phi_2$, and the
halo $\phi_3$, i.e.,
$\phi_{\rm Gal} = \phi_1 + \phi_2 + \phi_3$.
These three parts of the potential can be expressed in
cylindrical coordinates ($R,z$) as follows:
\begin{equation}
\label{eq:galcatic_potential}
\begin{split}  
  \phi_1 & = -\dfrac{M_1}{\left(R^2+z^2+{b_1}^2\right)^{1/2}}\ ,
  \\
  \phi_2 & = -\dfrac{M_2}{\left\{
           R^2+\left[a_2 + \left(z^2+{b_2}^2\right)^{1/2}
           \right]^2 
           \right\}^{1/2}}\ , 
  \\
  \phi_3 & = -\dfrac{M(r)}{r}-\dfrac{M_3}{1.02 a_3}
  \\
         & \times \left[-\dfrac{1.02}{1+(r/a_3)^{1.02}} + 
           \ln(1+(r/a_3)^{1.02})\right]^{r_\d}_{r}\ , 
\end{split}
\end{equation}
where
\begin{equation}
\label{eq:galcatic_potential_1}
\begin{split}  
r&=\left(R^2+z^2\right)^{1/2}\ ,
\\
M(r)&=-\dfrac{M_3(r/a_3)^{2.02}}{1+(r/a_3)^{1.02}}\ .
\end{split}
\end{equation}
The parameters for the above equations are
$M_1=1.41\times10^{10}~M_\odot$, $b_1=387~\pc$,
$M_2=8.56\times10^{10}~M_\odot$, $a_2=532~\pc$,
$b_2=250~\pc$, $M_3=10.7\times10^{10}~M_\odot$,
$a_3=1.2\times10^4~\pc$ and $r_\d=10^5~\pc$. The OC is
positioned on \response{a} circular orbits at a speed of
$220~\km~\s^{-1}$, specifically
$(v_{\rm R},v_\phi,v_{\rm z}) = (0,220,0)~\km~\s^{-1}$, at a
distance of $8.5~{\rm kpc}$ from the Galactic center.

\begin{figure} 
  \centering    
  \includegraphics[width=1.0\columnwidth]
  {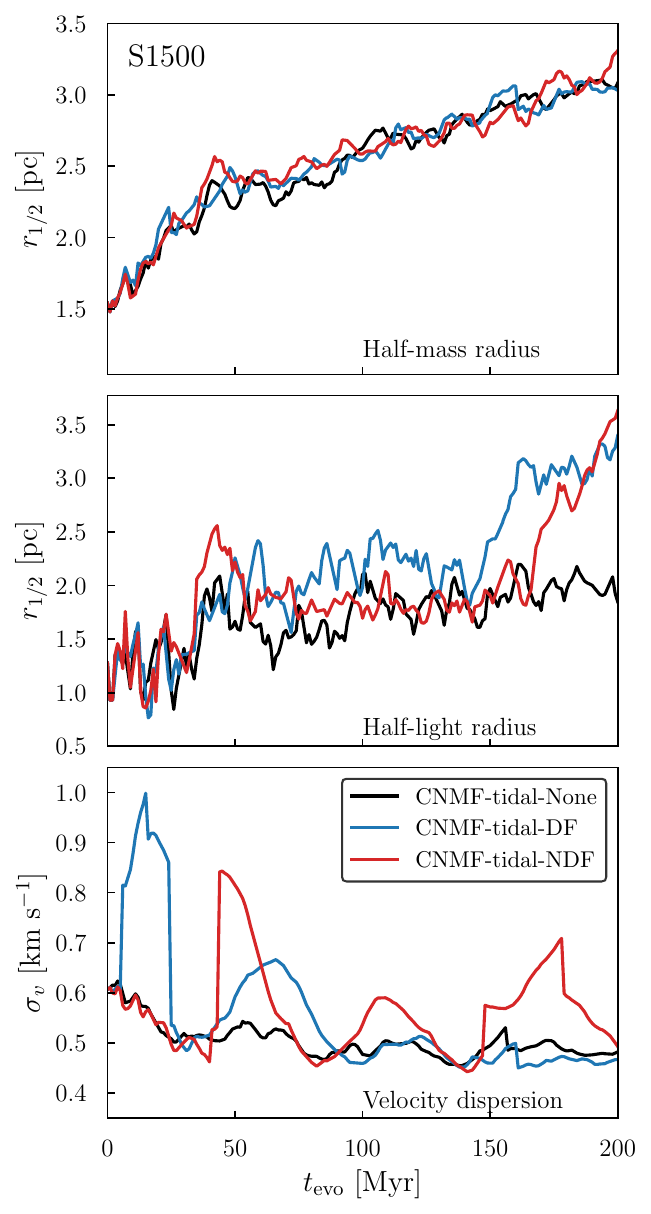}
  \caption{Similar to
    Figure~\ref{fig:CNMF-unseg}, but the results
    shown are the NDF scenario considered the Galaxy's
    potential.}
  \label{fig:CNMF-Radius-tidal-ndf}     
\end{figure}

  The impacts of the gas-star interactions in this
  galactic scenario are illustrated by
  Figure~\ref{fig:CNMF-Radius-tidal-ndf}. While the overall
  trend is consistent with Figure~\ref{fig:CNMF-unseg} that
  both the half-mass and half-light radii increase over
  time, the tidal fields dominate the dynamical evolution
  and generally dwarf the differences in the half-mass radii
  and velocity dispersions for the three gas-star
  interaction models. The half-light radii, which are more
  sensitive to massive stars, are greater for NDF and DF
  models compared to the gas-free case.
  \citet{2020A&A...640A..84D} studied the dynamical
  evolution of the tidal tails of OCs in detail and provided
  a semi-analytical model, concluding that such oscillations 
  in the velocity dispersions solely depend on the local 
  galactic \response{dynamical} properties,
  especially the orbital frequency and the epicycle
  frequency, rather than any cluster properties (e.g., the
  initial velocity dispersion or mass). 

It is expected that the tidal field could be one of the
significant effects of the evolution of internal dynamical
features within OC. However, this work focuses on the effect
of gas-star interactions. While incorporating the tidal
field would render the scenario more realistic, it also
complicates the physical mechanisms, making it challenging
to analyze the effects of different mechanisms
clearly. Therefore, the tidal field will not be considered
in subsequent scenarios, meaning the external Galaxy's
potential will be excluded. In future work, the influence of
both tidal fields and gas interactions on the dynamical
characteristics of stars or clusters should be included.

\section{Models Exploring Ambient Gas Properties}
\label{sec:result-var-model}

The Galaxy contains multiple phases of the ISM in
  its disk, whereas the CNM phase that we have explored in
\S\ref{sec:result-fiducial} only occupies a relatively small
volume fraction (typically a few percent, see e.g.,
\citealt{2001RvMP...73.1031F, DraineBook}).  We therefore
conduct various simulations to study the interactions
between OCs and other types of gases, including the WNM,
diffuse molecular regions (MD), dense and intermediate
molecular clouds, and even the mid-plane of \ac{AGN} disks
\citep{DraineBook, 2021ApJ...910...94C}. Properties of these
gases are summarized in
Table~\ref{tab:sims_list}. Note that all
  simulations in this section use the fiducial sample
  S1500.

\begin{figure} 
  \centering    
  \includegraphics[width=1.0\columnwidth]
  {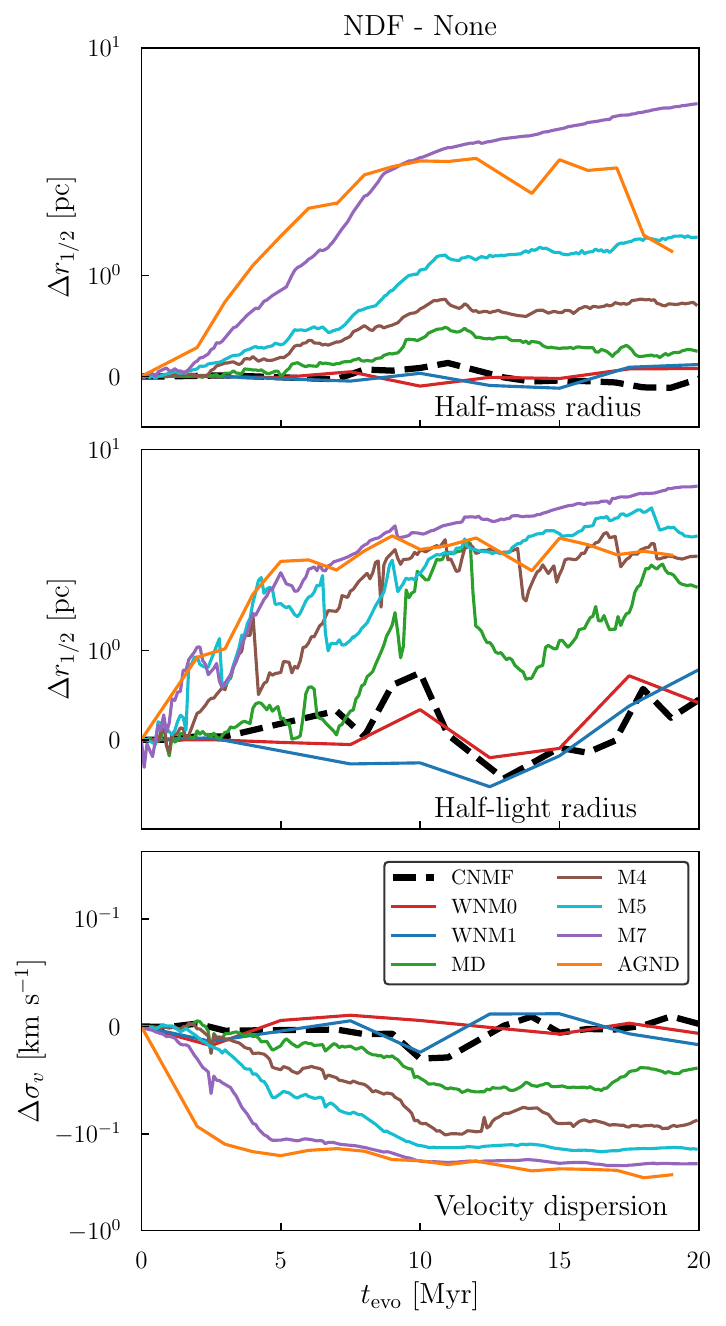}
  \caption{Similar to Figure~\ref{fig:Radius-NDF-DF-2}, but
    for various ambient gas densities without $v_{\rm c}$,
    showing only the first $20~\Myr$. Note that the cut-off
    \response{around $19~\Myr$} for AGND exists because no
    stars have negative \response{total} energy in the OC's
    center-of-mass frame.}
  \label{fig:Radius-NDF-DF}     
\end{figure}

\begin{figure} 
  \centering    
  \includegraphics[width=1.0\columnwidth]
  {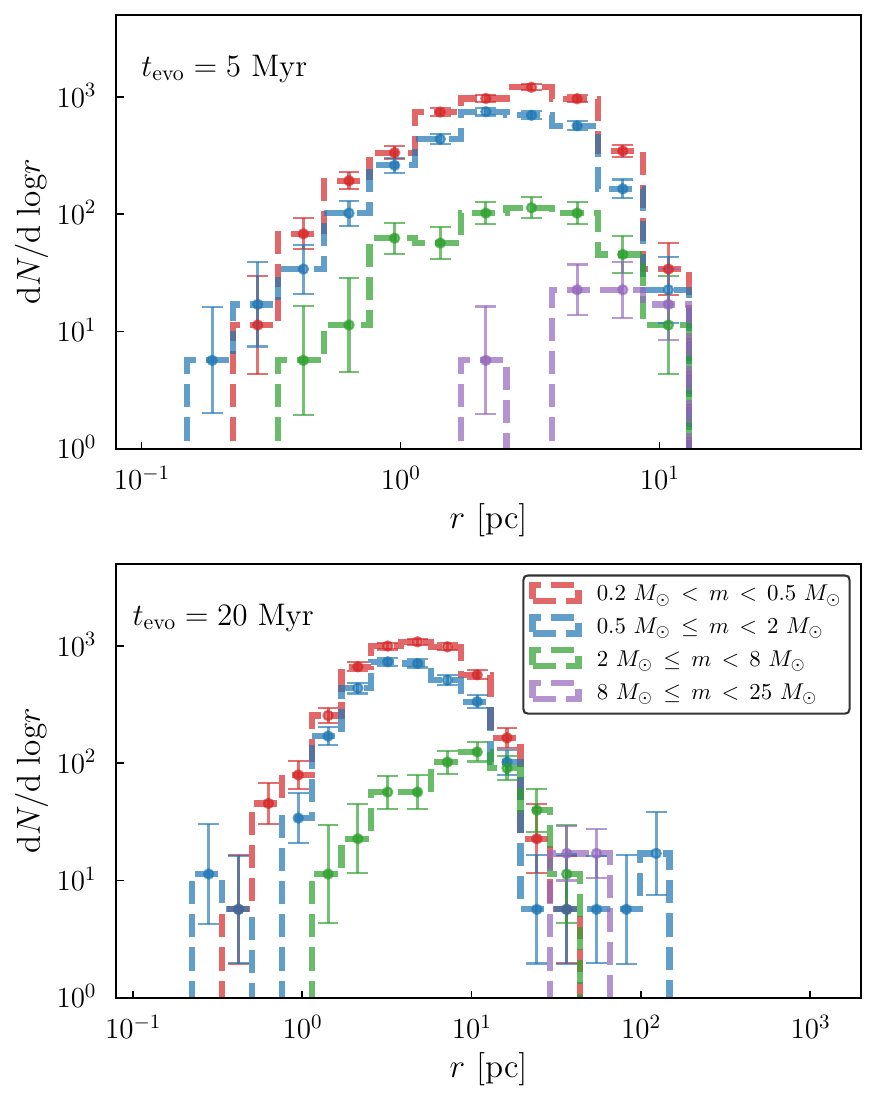}
  \caption{The distance distribution for
      different mass bins in the M5-NDF scenario 
      after $5~{\rm Myr}$ (upper panel) and $20~{\rm Myr}$ (lower panel) evolution. Due to
      their stronger winds, massive stars are subject to
      greater NDF than low-mass stars, resulting in a
      reduced mass segregation.}
  \label{fig:M5-mass-segregation-ndf}     
\end{figure}

\begin{figure} 
    \centering    
    \includegraphics[width=1.0\columnwidth]
    {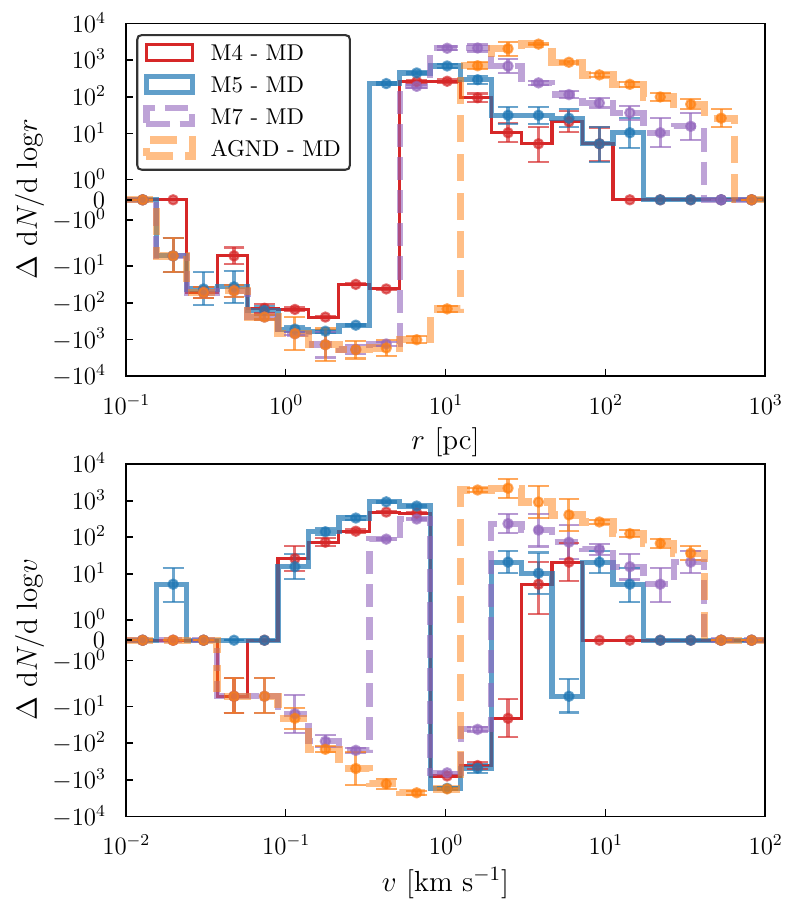}
    \caption{Similar to Figure~\ref{fig:CNMF-3-5-RV}, but the results shown are the difference in their distribution functions between the labeled models and the MD (diffuse molecular regions) model (see also Table~\ref{tab:sims_list}) for denser ambient gas without $v_{\rm c}$, influenced by NDF.}
    \label{fig:CNMF-more-RV}     
\end{figure}

\subsection{Distributions in the Configuration and \\
  Velocity Spaces}
\label{sec:result-var-size}

In a virialized self-gravitating system, stars staying at
larger distances from the cluster center generally move at
slower velocities. When the NDF and DF accelerations are
sufficiently low and gradual, their impact on the cluster
configurations is mostly adiabatic, and the system remains
virialized during the evolution.  When the ambient gas
density is raised to $3\times 10^8~m_p~\cm^{-3}$ in the
model AGND, however, the NDF acceleration for a solar-mass
star moving at $1~\km~\s^{-1}$ becomes
\response{$a_{\rm NDF}\approx 10^{-9}~\cm~\s^{-2}$}, which is
\response{about $10 \times$} the gravitational acceleration by a
$10^3~M_\odot$ cluster at $10~\pc$. A cluster evolving under
such strong NDF or DF accelerations no longer remains
adiabatic; the dispersal takes place directly within the
first \response{$3~\Myr$} resulting in a quickly expanding spatial
size and higher velocity dispersions
(Figure~\ref{fig:Radius-NDF-DF}). This phenomenon also
appears in the models M7 and M5 that stand for typical
molecular clouds, while the models for more
  diffuse gas (WNM0, WNM1) are qualitatively similar to the
  CNMF case.

Another effect that emerges in various ambient models is the
stratification of stars. The acceleration by NDF on stars
tend to increase their energy, which is similar to some
``buoyancy'' in self-gravitating OCs. As more massive stars
launch more powerful stellar winds (\S
\ref{sec:method-star-evo}), they tend to move quicker to the
``surfaces'' of OCs under NDF, and vice versa. We notice
that such ``buoyancy'' will not make the young OCs
``inversely segregated'' initially, as the timescales
required to fully ``inversely segregate'' the cluster
(\response{about $10^7-10^8~\yr$}) (see
e.g. Figure~\ref{fig:M5-mass-segregation-ndf}) is longer
than the time period of star formation (\response{$\lessapprox 10^7~\yr$})
\citep{1999ApJ...525..772P,2005ApJ...626L..49P}.

Since the stellar luminosity scales as $L\propto M^{3.5}$,
the half-light radii are considerably greater than half-mass
radii in NDF-affected clusters. For those models that have
dense ambients (e.g. M4, M5), the half-light radii
are \response{$\gtrapprox 3\times$} greater than the half-mass radii
after only \response{about $10~\Myr$}, indicating an obvious radial
stratification in terms of stellar types. When attempting to
infer the age of an observed OC by the observations in
astrometry and dynamics, one should take care of the
possible encounters with dense clouds, whose effect in
puffing up the OCs is similar to the intrinsic evaporation
of clusters. Statistics on the radial distributions of mass
and light in such clusters, therefore, may be helpful in
reducing this type of parameter degeneracies.

\subsection{Puff-up and Dispersal of Open Clusters in Gases}
\label{sec:result-var-puff}

According to eq.~\eqref{eq:method-anti-fric}, the magnitude
of NDF acceleration depends much more sensitively on
$\rho$ than on $T$. Figure~\ref{fig:CNMF-more-RV}
illustrates the general trend that the NDF effects increase
with denser ambient gas, raising the total \response{energy $E$} 
and easing the evaporation of cluster stars. The scaling
relation in eq.~\eqref{eq:method-anti-fric} is sub-linear
with respect to $\rho$
($\propto\rho^{1/2}$). Assuming a relatively
invariant velocity distribution over the same period of
time, one can infer the scaling relation
$\d E/\d t\propto\rho^{1/2}$, and subsequently
$\Delta |1/r|\propto\rho^{1/2}$. This pattern
is qualitatively seen in Figure~\ref{fig:Radius-NDF-DF}, yet
we note that such simple scaling no longer applies
quantitatively when there are significant increases in
velocity dispersions.

Models M5 and M7 show that OCs in relatively dense ambients
will quickly puff up or even disperse within
\response{$20~\Myr$}, a timescale that is comparable or
shorter than the crossing time of an OC over a dense
molecular cloud.  If one adopts the typical sizes
\response{(about $20~\pc$)} of dense molecular clouds
\response{($\rho\gtrapprox 10^4~m_p~\cm^{-3}$)}
\citep{1991ASIC..342..287C, 2007ARA&A..45..339B} with an OC
crossing time \response{$\approx 40~\Myr$} (assuming bulk
velocity \response{$\approx 0.5~\km~\s^{-1}$}), Model M5
indicates that a cluster should puff up to
\response{$r_{1/2} \gtrapprox 10~\pc$} after this
crossing. Similar results also occur in Model M7, even if
one limits the interaction time to the typical lifetime of
dense molecular clouds (about a few Myr). The encounter with
molecular clouds can lead to significant changes in its
speed due to NDF or DF, \response{scaled with respect to the
  data of solar-like stars},
\begin{equation}
  \label{eq:solar-like-v-ndf-df-mc}
  \begin{split}
    \Delta v_*^{\rm NDF} & = a_{\rm NDF}\Delta t \approx
                           0.8~\km~\s^{-1}\times 
          \left( \dfrac{v_\w}{400~\km~\s^{-1}} \right)^{1/4}
     \\
        &  \times
          \left(\dfrac{\rho}{3\times10^{6}~m_p~\cm^{-3}}
          \right)^{1/4} 
            \left( \dfrac{\Delta t}{10~\Myr} \right)^{1/2}
     \\
        &
          \times\left(\dfrac{\dot{m}_\w}
          {4\times10^{-13}~M_\odot~\yr^{-1}}\right)^{1/4}\ , 
     \\
    |\Delta v_*^{\rm DF}| & = |a_{\rm DF}\Delta t| \approx
                            16.5~\km~\s^{-1}\times 
          \left( \dfrac{\Delta t}{10~\Myr} \right)^{1/3}
     \\
        &\times
          \left(\dfrac{\rho}{3\times10^{6}~m_p~\cm^{-3}}
          \right)^{1/3}\ . 
  \end{split}
\end{equation}
\response{For AGB stars with massive outflows (see also 
eq.\eqref{eq:method-dv-pulse} in \S\ref{subsec:gas-star}), 
the increment in stellar velocity caused by NDF could 
raise further to $\Delta v_*^{\rm NDF} \approx 0.2~\km
~\s^{-1}$ after $10^5~\yr$.}

AGN disks are also considered as places where stars can
form, whose lifetimes are at the order of $10^0-10^3~\Myr$
\citep{2004MNRAS.351..169M, 2001ApJ...547...12M}. If a
$10^3~M_\odot$ OC forms and stays inside the AGN disk for
\response{$\gtrapprox 5~\Myr$}, it could evaporate almost \response{completely} before
it is no longer considered as a cluster anymore, even if the
destruction processes by orbital motion shears and tidal
forces are taken into account. In other words, the
likelihood that one can find OCs formed in AGN disks should
not be significant.

\subsection{Compact Objects and Binaries}
\label{sec:result-var-compact-bin}

\begin{figure} 
    \centering    
    \includegraphics[width=1.0\columnwidth]
    {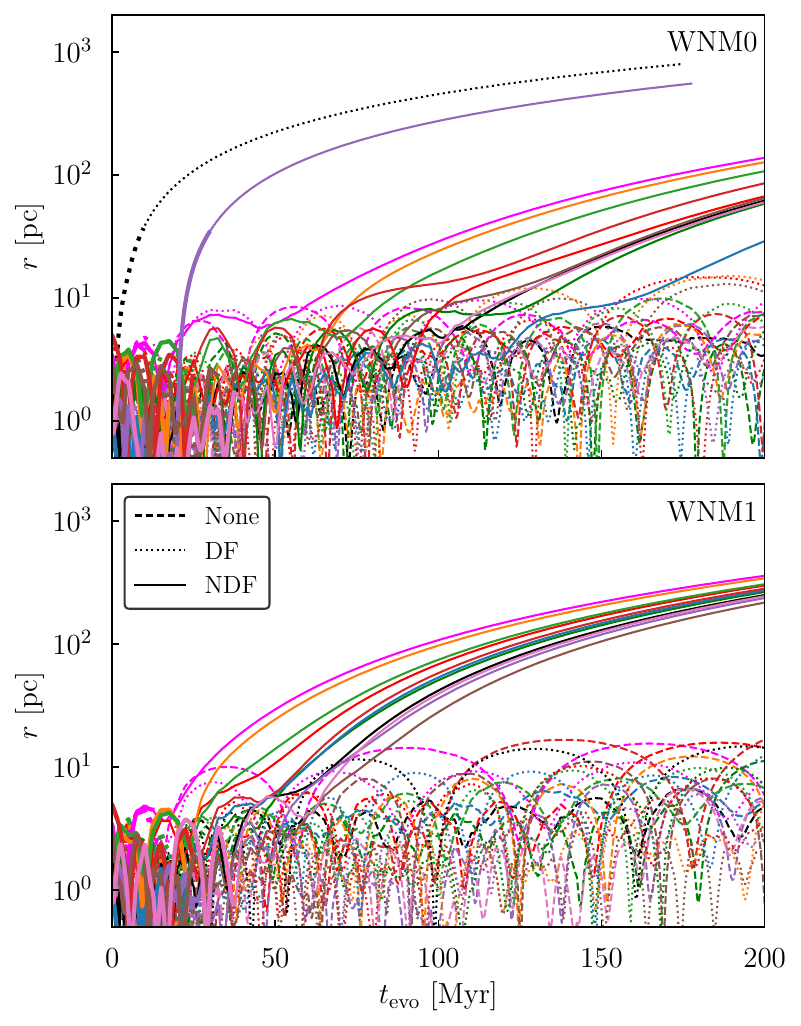}
    \caption{Similar to Figure~\ref{fig:track_CNM3_CNM5},
      but for the scenarios of WNM0 (upper panel) and WNM1
      (lower panel).}
    \label{fig:track_WNM0_WNM1}     
\end{figure}

\begin{figure} 
  \centering    
  \includegraphics[width=1.0\columnwidth]
  {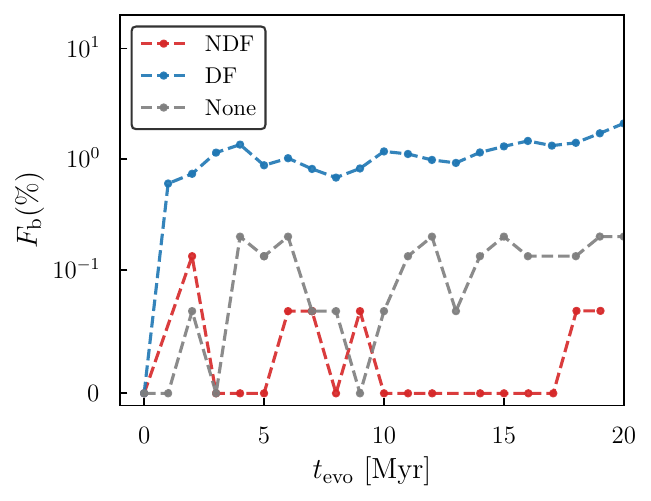}
  \caption{Similar to Figure~\ref{fig:binary_3e1-100}, but
    for the AGND scenario with an evolution time of
    $20~{\rm Myr}$.}
  \label{fig:binary-AGND}     
\end{figure}

As elaborated in \S \ref{sec:result-fiducial-bulk} and
\ref{sec:result-var-size}, higher stellar velocities and
lower ambient gas densities suppress the NDF
acceleration. In various models, the expulsion of NS via NDF
behaves accordingly. Figure~\ref{fig:track_WNM0_WNM1}
illustrates that, even for Model WNM1 whose ambient density
is merely $3~m_p~\cm^{-3}$, such expulsion mechanism is
still working, although the time required is considerably
longer. This effect is, nevertheless, still more perceptible
than the CNM5 case (see
Figure~\ref{fig:track_CNM3_CNM5}). For Model WNM0, the
expulsion mechanism weakened compared to WNM1, but all NSs
still tend to move away from the OC.

The binary fraction corresponding to various
models does not show any considerable changes compared to
the fiducial CNMF model, except for the AGND model with DF
only (Figure~\ref{fig:binary-AGND}). The binary
  fraction increased drastically in the first \response{$4~\Myr$}
  of evolution. Strong deceleration caused by DF shrinks
the separations between stars and reduces stellar
velocities, facilitating the formation of binaries. Several
studies have explored this
phenomenon. \citet{2013ApJ...774..144R,2016ApJ...827..111R}
proposed that gas-assisted inspirals efficiently merge
supermassive BH binaries within a Hubble time. The
\response{NDF} would nonetheless overturn this scenario,
widening the binary separations. \citet{2022ApJ...932..108W}
demonstrated in simulations of AGB star-outflowing pulsars
that a dense and slow outflow exerts a positive torque on
the binary, causing \response{$\gtrapprox 10\%$} orbit
expansion. Extrapolating these mechanisms to the OCs
immersed in dense gas can cause a severe inhibition of
binary occurrence rates.

\section{Discussion and Summary}
\label{sec:summary}
\response{In this paper, we study the evolution of OCs immersed in the ISM.} 
Overall, the introduction of NDF makes the OC puffier
and eases the evaporation of cluster stars, resulting in
reduced velocity dispersions and increased half-mass and
half-light radii. This effect scales sub-linearly with the
ambient density and is mostly irrelevant to the ambient
temperature as the ram pressures on both sides 
\response{of the contact discontinuity} dominate the
gas-star interactions. NS with powerful winds are expelled
from the cluster due to the intense NDF effect, while WD
stars undergo conventional evaporation due to the absence of
\response{a WD-driven wind}.

\subsection{Impacts of the NDF Effects on Cluster Evolution}
\label{sec:summary-impact}
  
While gas DF has garnered most of the attention, NDF has
been relatively overlooked. Opposite to DF, NDF accelerates
stars along their motion direction. Such difference in the
effects between NDF versus DF can ``flip'' the signs of
multiple physical mechanisms. For instance, one may expect
that the OC evaporation is suppressed or even inhibited by
DF when encountered with gas, while the actual scenario
could be right the opposite due to NDF. In the densest gases
including dark molecular clouds, AGN disks, and AGN tori
(whose average column density is as high as
\response{ about $10^{24}~{\rm cm}^{-2}$}; see, e.g.,
\citealt{2021A&A...650A..57Z}), dispersal of OCs can be
almost instant compared to their dynamical lifetimes as
clusters.  Interpretations about observations on cluster
kinetics should take the possibility of NDF into account.
\citet{2009MNRAS.397..488P} argued that massive
  stellar clusters over a certain critical mass can act as
  cloud condensation nuclei and accrete gas from the ambient
  medium. Such critical mass is required for the cluster to
  accrete gas and induce density instability in the ISM
  roughly satisfies
  $M_{\rm cl}/T_{\rm g}b_{\rm cl} >
  3.33\times10^3~M_\odot~{\rm K}^{-1}~\pc^{-1}$ (where
  $M_{\rm cl}$ is the cluster mass, $T_{\rm g}$ is the gas
  temperature, and $b_{\rm cl}$ is the Plummer parameter;
  see also \citealt{1911MNRAS..71..460P}). Admittedly,
massive clusters can suppress the NDF effect by accreting
from the \response{ambient gas}, which is also beyond the scope of the
current study and will be addressed in future works. In the
scenarios considered in this work, nevertheless, the cluster
mass is insufficient to initiate this type of accretion.  

Observation missions like Gaia \citep{GAIA2016} provide
abundant astrometric data for stars in our galaxy,
facilitating the study of OCs' dynamical evolution
\citep[e.g.,][]{2023JApA...44...71M}. Because of the radial
stratification in stellar masses (\S
\ref{sec:result-var-size}), one can prospectively tell the
effects of gas-star interactions by measuring the half-mass
and half-light radii, as well as the velocity
dispersions. The combination of high precision photometry
with astrometry can further help the researchers to find out
the radial distribution functions of different stellar
types. These measurements can be a direct characterization
of gas-star interaction history that are potentially
important for identifying an OC's ``invasion'' into dense
gases during its evolution history.

\subsection{Neutron Star Depletion}

Due to substantial outflows, NSs gradually drift away from
their associated OC under the influence of NDF in all
simulations. Conversely, WD stars influenced by NDF
experience a pulse momentum injection from the main sequence
to the AGB stage and subsequently to the WD stage.  However,
upon transitioning into the WD phase, WDs decelerate due to
DF as they lack outflows. This mechanism ensures that part
of WDs remain confined within the OC and experience
conventional evaporation. \response{If the kick received by a WD is
strong enough, such as through an asymmetric mass loss
process \citep{2020ApJ...901L..16F}, it is more likely to
be expelled from the OC (see in
Figure~\ref{fig:wd_track_3e1-100n}).}

It is commonly assumed that kicks by asymmetric supernovae
drive NS to high velocities relative to the cluster centers,
and eventually expel them from OCs.
\citet{2004cetd.conf..276L} reviewed various physical
mechanisms leading to kicks, and \citet{2017MNRAS.469.1510B}
suggested that NSs are generally scattered away from SgrA*
due to such kicks. \citet{2015MNRAS.449L.100C} indicated
that the natal kick of NSs can alter the cluster's lifetime
by almost a factor of \response{4}. Admittedly, the kicks serve
as an effective expulsion mechanism of NS.  However, kicks
only entail a one-time momentum injection, whose intensity
and direction of momentum injection is stochastic and can
well result in the NS's deceleration. The NDF effects are,
in contrast, a much more steady and robust mechanism that
almost ensures the expulsion without gravitational
scatterings. Such effects can further lead to the reduction
of NS binaries and mergers within OCs, which is potentially
relevant to the spatial distribution of NS-related mergers
and gravitational wave events.

\subsection{Future Works}
\label{sec:summary-future}

Due to the limitations in physical modeling and computation,
this work does have some caveats and issues that should be
addressed in future works. For example, simplistic isotropic
stellar wind models are adopted for each type of star, which
is suitable only for a fraction of NS. Many NS--especially
those with accretion disks--may launch directed or bipolar
outflows, which could affect the intensity and direction of
the NDF forces \citep{2020MNRAS.494.2327L}.  The WD model in
this study lacks wind and is susceptible to DF only. While
stand-alone WD stars may not have outflows, accretion from
their companions may cause disk winds and jets. This study
ignores the time lag in forming stars within the OC and does
not consider the interaction between star and star-forming
gas. However, the expulsion of residual star-forming gas
does not occur instantaneously after the completion of the
star formation process in realistic scenario.  Observations
reveal that star formation efficiencies within OCs vary
widely, ranging from several percent to 30 percent for dense
clumps within molecular clouds
\citep{2003ARA&A..41...57L,2009ApJ...705..468H,2016AJ....151....5M}, and from
0.1 percent to a few percent for their associated giant
molecular clouds
\citep{2009ApJS..181..321E,2011ApJ...729..133M}. This
implies that newly formed or forming stars (such as
protostars and pre-MS stars, which have outflows) would
interact with star-forming gas, resulting in the influence
of NDF on the initial phase space distribution of stars
within young OC. Moreover, the stellar winds from these
stars may blow unprocessed gas out of the cluster,
inhibiting the ongoing star formation process.  Refined
models in the future should take the complexities of stellar
outflows and the star formation process into account.
Besides, this work neglects the interactions
  between the outflows of stars in binary systems and
  between binary systems and the outflows of single
  stars. These effects will be investigated in future work
  using more refined hydrodynamic simulations.

Although our study considers the highly dense scenario of an
AGN disk (AGND scenario), we have omitted multiple physical
conditions relevant to AGN in order to isolate and emphasize
the impact by NDF.  Incorporating the shear effect is
crucial for understanding stellar formation and evolution
within AGN disks.  However, this topic is specific to disks
and beyond the scope of this paper, which should be
addressed in a separate study. Future studies on AGN disks
will require additional considerations, such as orbital
motion and vertical stratification of gas.  The limited
thickness of AGN disks differs substantially from the
assumed uniform gas ambient in this paper, especially
considering the vertical density gradients. Furthermore,
isolated stars would experience the DF effects in the
highly dense environment of a galaxy's circumnuclear
regions. Future works shall also explore these possibilities
of NDF-affected stars.

In addition, this work focuses on relatively small and
sparse OCs with \response{about $10^3$} stars. In high-density
environments, particularly in AGN disks, super star clusters
are often present instead of OCs, whose star formation
efficiencies \response{may be typically} high. Investigating larger
ensembles of stars such as globular clusters can impose
additional challenges. Globular clusters \response{contain about $10^6$} or
even more stars, while each of the stars might still be
dynamically important.  Accurate treatments of frequent
close encounters are beyond the capacity of ordinary orbital
integrators with tree-based or particle-mesh-based gravity
solvers, and special algorithms are required for sufficient
accuracy. \response{During close encounters, shock structures caused
by gas-star interactions may collide},
resulting in highly complex gas morphologies and interaction
patterns. Understanding how such complex physical scenarios
affect the dynamical evolution of stars within globular
clusters could be explored once proper algorithms are
prepared for complicated computations. Even without proper
treatments of frequent close encounters in more massive
clusters, the dynamical evolution of compact objects (NS,
WD, and BH) are still qualitatively feasible within the
current framework for globular clusters and super star
clusters. Detailed discussions are nontheless beyond the
scope of the current paper focusing on OCs, and a subsequent
paper is being composed specifically addressing this issue.

\begin{acknowledgments}
  L. C. Ho was supported by the National Science Foundation
  of China (11991052, 12233001), the National Key R\&D
  Program of China (2022YFF0503401), and the China Manned
  Space Project (CMS-CSST-2021-A04, CMS-CSST-2021-A06).
  X. Fu thanks the support of the National Natural Science
  Foundation of China (NSFC) No. 12203100 and the China
  Manned Space Project with NO. CMS-CSST-2021-A08.  M. Liu
  and L. Wang appreciates the computational resources
  provided by the Kavli Institute of Astronomy and
  Astrophysics at Peking University.  We also thank our
  colleagues Renyue Cen, Kohei Inayoshi, Fangzhou Jiang, and
  Meng Sun for helpful discussions and suggestions.
\end{acknowledgments}

\bibliographystyle{aasjournal}
\bibliography{NDF-OC}{}

\end{document}